\documentclass[aps,pra,twocolumn,showpacs,longbibliography]{revtex4-1}  % for review and submission
\usepackage{graphicx}  % needed for figures
\usepackage{dcolumn}   % needed for some tables
\usepackage{bm}        % for math
\usepackage{amsmath, amsthm, amssymb}
\usepackage{bbold} % to generate identity 1
\usepackage{mathrsfs}
\usepackage{array}
\usepackage[toc]{appendix}
\usepackage[dvipsnames,usenames]{color}
\usepackage{soul} % to use \st for strikeout 
\usepackage{ulem} % to use \sout for strikeout to include equations
\usepackage{hyperref}
\hypersetup{colorlinks=true, linkcolor=RoyalBlue, citecolor=RoyalBlue, urlcolor=RoyalBlue}

\def\ave#1{\langle #1\rangle}

\newcommand{\bs}[1]{\boldsymbol{#1}}

\newcommand{\iv}{\mathbf{i}}

\newcommand{\qv}{\mathbf{q}}

\newcommand{\vell}[1]{\bs{\ell}}

\begin{document}

\normalem

\title{{Increasing superconducting $T_c$ by layering in the attractive Hubbard model}}
\author{Rodrigo A. Fontenele}
\author{Natanael C. Costa}
\author{Thereza Paiva}
\author{Raimundo R. dos Santos}
\affiliation{Instituto de F\'isica, Universidade Federal do Rio de
Janeiro Cx.P. 68.528, 21941-972 Rio de Janeiro RJ, Brazil}
\begin{abstract}
The attractive Hubbard model has become a model readily realizable with ultracold atoms on optical lattices.
However, the superconducting (superfluid) critical temperatures, $T_c$'s, are still somewhat smaller than the lowest temperatures achieved in experiments. 
Here we consider two possible routes, generically called layering, to increase $T_c$: a bilayer and a simple cubic lattice, both with tunable hopping, $t_z$, between  attractive Hubbard planes.
We have performed minus-sign--free determinant quantum Monte Carlo simulations to calculate response functions such as pairing correlation functions, uniform spin susceptibility, and double occupancy, through which we map out some physical properties. 
We have found that by a judicious choice of fillings and intensity of on-site attraction, a bilayer can exhibit $T_c$'s between 1.5 and 1.7 times those of the single layer; for the simple-cubic lattice the enhancement can be 30\% larger than the maximum for the single layer.
We also check the accuracy of both a BCS-like estimate for $T_c$ in the attractive Hubbard model, as well as of an upper bound for $T_c$ based on the superfluid density.  
\end{abstract}
\date{Version 7.3 -- \today}
\maketitle

%%%%%%%%%%%%%%%%%%%%%%%%%%%%%%%%%%%%%%%%%%%%%%%%%%%%%%%%%%%%%%%%%%
\section{Introduction}
%%%%%%%%%%%%%%%%%%%%%%%%%%%%%%%%%%%%%%%%%%%%%%%%%%%%%%%%%%%%%%%%%%

The attractive Hubbard model (AHM) describes the dynamics of fermions moving in a single band (nearest neighbor hopping integral $t$) subject to an on-site interaction, $U<0$, which favors the formation of local pairs \cite{Micnas90}. 
This model has been useful to study several phenomenological aspects of superconductivity. 
For instance, it contemplates the existence of a temperature scale, $T_p\gtrsim T_c$, with $T_c$ being the critical temperature for superconductivity, associated with a suppression of the uniform spin susceptibility, $\chi_{s}$, signalling the opening of a spin gap \cite{Randeria92,Bucher93,dosSantos94,Paiva10,Fontenele22}. 
This connects directly with pseudo-gap phenomena in high-temperature cuprate superconductors \cite{Wilson01}. 
Another important feature of the AHM is the possibility of interpolating between the weak coupling limit, displaying BCS behavior, with long coherence length pairs, and the strong coupling regime, with tightly-bound pairs and short coherence length undergoing Bose-Einstein condensation \cite{Micnas90,Randeria95a,Chen05,Randeria14, Fontenele22}.

With the recent advances in optical lattice experiments (OLE) involving ultracold fermionic atoms \cite{chin10,Jaksch05,Bloch08,Esslinger10,McKay11,Mitra18,kozik10}, the AHM has found an immediate experimental realization, in which there is a fine control of the interaction strength through an external magnetic field. 
The advent of the quantum gas microscope \cite{Bakr09} paved the way to visualise the atomic distribution on the lattice and draw quantitative conclusions, such as  correlation functions for the AHM on a square optical lattice \cite{Mitra18,Gall20,Chan20,Hartke23}.
Although OLE constitute an important setup for studying strongly correlated phenomena, the current lowest temperatures achieved 
are still higher than theoretical predictions for $T_c$. More specifically, recent extensive Quantum Monte Carlo (QMC) simulations  \cite{Fontenele22} mapped the square lattice phase diagram, $T_c (\langle n\rangle,U)$, with $\langle n\rangle$ being the band-filling: a maximum $T_c\approx 0.15 t/k_B$ occurs around $\langle n\rangle \approx 0.87$  \cite{Fontenele22}, where $k_B$ is the Boltzmann constant, which we set to unity from now on. This corresponds to a few tens of nanokelvin, which is still about a third of the current lowest temperatures experimentally accessible in OLE \cite{Hartke23,Mitra18}. Hence, it is important to investigate scenarios which could lead to higher $T_{c}$'s, closer to an experimentally accessible  range.

As a rough guide to this quest, we recall that the BCS behavior of the AHM at weak coupling  allows us to expect \cite{Micnas90}
\begin{equation}
    T_{c}^\text{BCS}= \widetilde{W} \exp{\left(-1/D(0)|U|\right)}~,
    \label{eq:Tc_BCS}
\end{equation}
where $\widetilde{W}$ is proportional to the bandwidth, $W$, 
and $D(0)$ is the density of states (DOS) at the Fermi level; throughout this paper we take $\widetilde{W} = W$ since the proportionality constant is immaterial to pinpoint the maxima critical temperatures. Therefore, one possibility to enhance $T_c$ is to increase $D(0)$, while keeping fixed both the fermionic density and $U$. 
For two-dimensional systems this may be achieved by adding a second layer and adequately tuning the hopping, $t_z$, between the layers; see below.
Note, however, that in two dimensions this is only applicable to the doped case, since the degeneracy between charge-density wave and superconductivity suppresses $T_c$ to zero at half-filling, by virtue of the Mermin-Wagner theorem.
Physically, the presence of a second layer offers another hopping channel, thus allowing pre-formed pairs to be less scattered throughout the lattice. 
We note that this degeneracy at half filling may be broken by considering a band insulating bilayer with attractive on-site interactions, thus allowing for a finite $T_c$  \cite{Prasad14, Prasad22}.

The possibility of a bilayer inducing a behavior different from the monolayer has been considered in the \emph{repulsive} Fermi-Hubbard model: it was found that interplane antiferromagnetic correlations tend to enhance $d$-wave pairing \cite{Scalettar94,dosSantos95}. 

Similar arguments apply to a simple cubic lattice, where a tunable $t_z$ could benefit from the fact that $T_c\neq0$ for  $t_z=t$ even at half filling \cite{dosSantos94}. 
By varying $t_z$ between 0 and 1, we follow a dimensional crossover between two- and three dimensions.  
Recent OLE on Bose-Einstein condensates with Rubidium-87 atoms allowed the study of dynamical properties crossing over between 2D and 3D behavior \cite{Zheng23}; this system can be modeled by a  Bose-Hubbard model with anisotropic hopping. At this point, we note that the minimum experimentally accessible temperature in OLE is dictated by the cooling protocols used, e.g. adiabatic cooling and sympathetic cooling \cite{Lewenstein07, Ketterle08, McKay11}, and not by the dimensionality of the optical trap (2D or 3D) \cite{Kohl05, Kohl06, Poulsen014, Tarruell18}.
Nonetheless, given that in strong-coupling the half filled (attractive or repulsive) Hubbard model can be mapped onto a Heisenberg model with antiferromagnetic exchange coupling $J=4t^{2}/|U|$, a mean-field argument leads to a critical temperature $T_c \sim zJ/2$, with $z=2d$ for hypercubic lattices; hence, $T_c$ for superconductivity is expected to increase with lattice dimensionality, $d$. Indeed, QMC simulations on the three-dimensional anisotropic repulsive Hubbard model at half-filling have found that a finite hopping between layers can enhance the magnetic structure factor relative to the $2$D case, and produce a slight increase in the critical temperature \cite{Ibarra20}. In addition, a perturbative cluster approach to the AHM suggests that the order parameter is enhanced relative to the two-dimensional case \cite{Ogawa15}.

In view of these possibilities, a systematic study of $T_c$ for the anisotropic AHM both in the bilayer and in three dimensions is of interest. 
With this in mind, here we use Determinant Quantum Monte Carlo (DQMC) simulations to investigate the AHM on these geometries.

The layout of the paper is as follows. In Section \ref{sec:HQMC} we discuss the model and highlight the main aspects of DQMC, including the quantities used to probe the physical properties of the system. 
In Section \ref{sec:bilayer} we present estimates for the critical temperature as a function of interlayer hopping on a square bilayer, followed by a discussion on the pairing temperature scale. 
In Section \ref{sec:cubic}, we consider the simple cubic lattice with variable interlayer hopping, $t_z$. In both cases we analyze the superconducting properties at different band fillings and  strengths of the on-site attraction. Finally, in Section \ref{sec:conc} we present our conclusions.

%%%%%%%%%%%%%%%%%%%%%%%%%%%%%%%%%%%%%%%%%%%%%%%%%%%%%%%%%%%%%%%%%%
\section{Model and Methodology}
\label{sec:HQMC}
%%%%%%%%%%%%%%%%%%%%%%%%%%%%%%%%%%%%%%%%%%%%%%%%%%%%%%%%%%%%%%%%%%

%%%%%%%%%%%%%%%%%%% Fig 1 %%%%%%%%%%%%%%%%%%%%%%%%%%%%%%%%%%%%%%%%%%%%%%%
\begin{figure}[t]
%  \centering
  \includegraphics[scale=0.6]{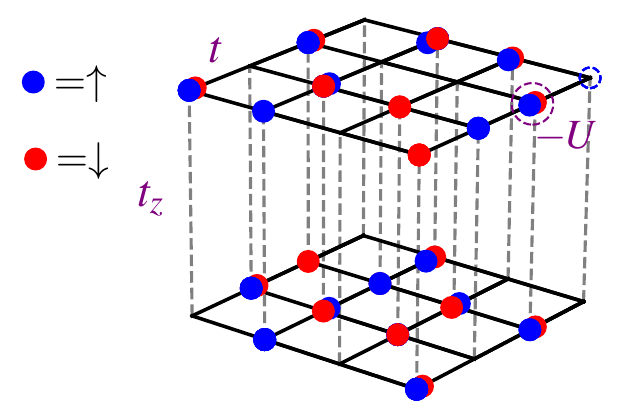}
   \caption{(Color online) 
   Fermionic atoms trapped in a bilayer. Each atom can be in either an $|\!\uparrow\rangle$ state (blue) or a $|\!\downarrow\rangle$ state (red). An atom may hop to its neighboring site at a rate $t$ within a plane, and $t_{z}$ between planes. 
   When two atoms in opposite spin states occupy the same site, the  energy is lowered by $|U|$.
   }
 \label{fig:Layers_rep}
\end{figure}

%%%%%%%%%%%%%%%%%%%%%%%%%%%%%%%%%%%%%%%%%%%%%%%%%%%%%%%%%%%%%%%%%%
We write the anisotropic attractive Hubbard Hamiltonian as,
\begin{align}
\mathcal{H} &= -t\sum_{\langle \mathbf{i}, \mathbf{j}\rangle_{\parallel},\sigma}\left(c_{\mathbf{i}\sigma}^{\dagger}c_{\mathbf{j}\sigma}^{\phantom{\dagger}} + \mbox{h.c.}\right) - t_{z}\sum_{\langle \mathbf{i}, \mathbf{j}\rangle_{\perp},\sigma}\left(c_{\mathbf{i}\sigma}^{\dagger}c_{\mathbf{j}\sigma}^{\phantom{\dagger}} + \mbox{h.c.}\right) \nonumber \\
&~~ -\mu\sum_{\mathbf{i}\sigma}n_{\mathbf{i}\sigma} - |U|\sum_{\mathbf{i}}\left(n_{\mathbf{i}\uparrow}-1/2 \right)\left(n_{\mathbf{i}\downarrow}-1/2 \right),
\label{eq:3D_Hamilt}
\end{align}
where $\langle \mathbf{i}, \mathbf{j}\rangle_{\parallel}$ represents nearest-neighbor sites within a layer, and $t$ is the intralayer hopping amplitude; $t_{z}$ is the hopping energy between neighboring sites in adjacent layers $\langle \mathbf{i}, \mathbf{j}\rangle_{\perp}$ (see Fig. \ref{fig:Layers_rep}); $\mu$ is the chemical potential controlling the band filling, and $|U|$ is the on-site attractive interaction between fermions of opposite spin. We consider unit lattice spacing, and $t$ sets the energy scale.

The physical quantities of interest are obtained through DQMC simulations \cite{Blankenbecler81,hirsch83,hirsch85,white89,dosSantos03b}. This is an unbiased numerical approach based on an auxiliary-field decomposition of the interaction, which maps onto a quadratic form of free fermions coupled to bosonic degrees of freedom, $\mathcal{S}(\iv,\tau) = \pm 1$, in both spatial and imaginary time coordinates. This method is based on a separation of the non-commuting parts of the Hamiltonian by means of the Trotter-Suzuki decomposition, i.e.
\begin{align}
    	\mathcal{Z} &= \mathrm{Tr}\,e^{-\beta\widehat{\mathcal{H}}}
	= \mathrm{Tr}\,[(e^{-\Delta\tau(\widehat{\mathcal{H}}_{0} + \widehat{\mathcal{H}}_{\rm
U})})^{M}]
\thickapprox \nonumber \\
& \thickapprox \mathrm{Tr}\,[e^{-\Delta\tau\widehat{\mathcal{H}}_{0}}e^{-\Delta\tau\widehat{\mathcal{H}}_{\rm
U}}e^{-\Delta\tau\widehat{\mathcal{H}}_{0}}e^{-\Delta\tau\widehat{\mathcal{H}}_{\rm
U}}\cdots], 
\end{align}
where $\widehat{\mathcal{H}}_{0}$ contains the terms quadratic in fermion creation and destruction operators,  and $\widehat{\mathcal{H}}_{\rm U}$ contains the quartic terms. 
We take $\beta = M \Delta\tau$ , with $\Delta\tau$ being the grid of the imaginary-time coordinate axis. This decomposition leads to an error proportional to $(\Delta\tau)^{2}$, which can be systematically reduced as $\Delta\tau\to 0$. Throughout this work, we choose $\Delta\tau \leq 0.1$ (depending on the temperature), which is small enough to lead to systematic errors smaller than the statistical ones (from the Monte Carlo sampling). Finally, for the AHM the discrete Hubbard-Stratonovich transformation dealing with the quartic terms in $\widehat{\mathcal{H}}_{\rm U}$ leads to sign-free simulations\,\cite{hirsch83,hirsch85,white89,dosSantos03b}.

In order to probe the emergence of superconductivity, we analyze the $s$-wave pair correlation functions,
\begin{equation}
C_{\mathbf{i}\mathbf{j}}^\Delta \equiv 
	\frac{1}{2}\langle b_{\mathbf{i}}^{\dagger} b_{\mathbf{j}}^{\phantom{\dagger}} 
	+ \text{H.c.}\rangle,
\label{eq:Cbibj}
\end{equation}
with $b_{\mathbf{i}}^{\dagger} \equiv c_{\mathbf{i}\uparrow}^{\dagger} c_{\mathbf{i}\downarrow}^{\dagger}$ ($b_{\mathbf{i}}^{\phantom{\dagger}} \equiv c_{\mathbf{i}\downarrow}^{\phantom{\dagger}}c_{\mathbf{i}\uparrow}^{\phantom{\dagger}}$) corresponding to  creation (annihilation) of a pair of fermions at a given site $\mathbf{i}$. Further, the Fourier transform of $C_{\mathbf{i}\mathbf{j}}^\Delta$ at $\mathbf{q}=0$ defines the \textit{s}-wave pair structure factor,
\begin{equation}
	P_{s}(\mathbf{q}=0)  = \frac{1}{N}\sum_{\mathbf{i}, \mathbf{j}}  C_{\mathbf{i}\mathbf{j}}^\Delta
\label{eq:Ps-def}
\end{equation}
with $N=L\times L\times L_{z}$ being the number of sites of the lattice, where $L$ is the linear dimension of the $2$D layers. 
For the bilayer, $L_{z}=2$, and for the cubic lattice, $L_{z} = L$;  periodic boundary conditions (PBC) are assumed in both cases.
One should keep in mind that in Eq.\,\eqref{eq:Ps-def} the dependence of $P_s$ on $L$, $U$, $t_z$, $\ave{n}$, and $\beta$ has been omitted, to simplify the notation.  

We estimate the inverse critical temperature, $\beta_c\equiv 1/T_c$, through the correlation ratio,
\begin{align}
    R_{c}(L,\beta) = 1 - \frac{P_{s}(\mathbf{q} + \delta\mathbf{q})}{P_{s}(\mathbf{q})}~,
    \label{eq:Rc_def}
\end{align}
where $\mathbf{q} = 0$, and $\delta\mathbf{q} = 2\pi/L$.
As highlighted in the Appendix, this quantity is a renormalisation-group invariant at the critical point\,\cite{kaul15,sato18,darmawan18}; that is, crossings of curves for $R_{c}(L,\beta)$ for different system sizes provide estimates for $\beta_c$, which improve as $L$ increases.

For our purposes here, the magnetic properties are probed by the uniform susceptibility. This is obtained through the fluctuation-dissipation theorem, which leads to $\chi_s = \beta \langle S^2 \rangle$, with $\langle S^2 \rangle$ being the uniform spin structure factor,
\begin{equation}
\langle S^2 \rangle =\frac{1}{N}\sum_{\mathbf{i}\mathbf{j}} \ave{\mathbf{S}_{\mathbf{i}}\cdot\mathbf{S}_{\mathbf{j}}},
\label{eq:magsusc}
\end{equation}
where $\mathbf{S}_{\mathbf{i}}\equiv (1/2)\mathbf{m}_{\mathbf{i}}$, with the components of the magnetization operator being
\begin{subequations}
\label{eq:mags}
\begin{eqnarray}
m_{\mathbf{i}}^x&\equiv&
c_{\mathbf{i}\uparrow}^\dagger c_{\mathbf{i}\downarrow}^{\phantom{\dagger}}+ c_{\mathbf{i}\downarrow}^\dagger c_{\mathbf{i}\uparrow}^{\phantom{\dagger}},
\label{eq:mx}\\
m_{\mathbf{i}}^y&\equiv&
-i\left(c_{\mathbf{i}\uparrow}^\dagger c_{\mathbf{i}\downarrow}^{\phantom{\dagger}}- c_{\mathbf{i}\downarrow}^\dagger c_{\mathbf{i}\uparrow}^{\phantom{\dagger}}\right),
\label{eq:my}\\
m_{\mathbf{i}}^z&\equiv&n_{\mathbf{i}\uparrow}-n_{\mathbf{i}\downarrow}.
\label{eq:mz}
\end{eqnarray}
\end{subequations}

%% fig 2 %%%%%%%%%%%%%%%%%%%%%%%%%%%%%%%%%%%%
\begin{figure}[t]
  \centering
  \includegraphics[scale=0.3]{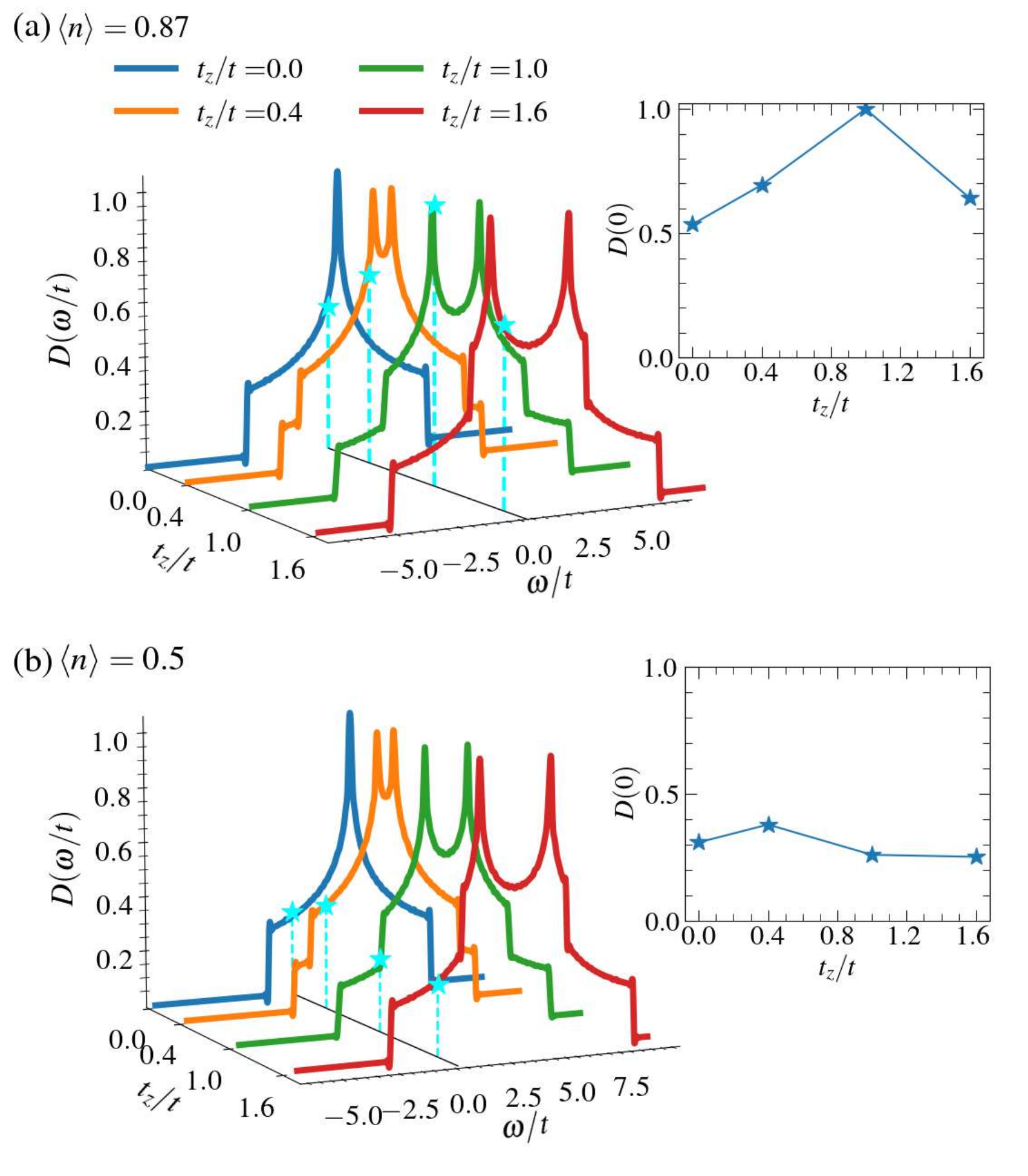}
   \caption{Density of states (DOS) for the non-interacting bilayer as a function of energy, $\omega = \varepsilon - \mu(t_{z})$, and interlayer hopping, $t_z/t$; the Fermi energy for a given electronic density is located at the origin, for (a) $\langle n \rangle = 0.87$ and (b) $\langle n \rangle = 0.5$. 
   The cyan stars in the main panels mark $D(0)$, as plotted in the insets as functions of $t_z/t$. 
   }
 \label{fig:DOS_n1.0}
\end{figure}

%%%%%%%%%%%%%%%%%%%%%%%%%%%%%%%%%%%%%%%%%%%%%%%
In what follows, we separate the discussion in two parts. 
First, we consider a bilayer, in which case our simulations were carried out for $L \leq 16$. 
Subsequently, we study the actual dimensional crossover in cubic systems with $L \leq 10$. 
Typically our data have been obtained after $2 - 4\times 10^{4}$ warming-up steps, followed by $8-16 \times 10^{4}$ sweeps for measurements, depending on the temperature, interaction strength, and electronic density. We have performed $10$ to $20$ realizations for each set of parameters to ensure the best accuracy for our results.

\section{The bilayer}
\label{sec:bilayer}

\subsection{Critical temperature}
\label{ssec:biTc}

We start the discussion with the non-interacting bilayer. 
The single-particle energies are given by 
\begin{equation}
    \varepsilon_\mathbf{k} = -2t\left[ \cos k_{x} + \cos k_{y}\right] \pm t_{z}~, 
    \label{eq:dispersion_bil}
\end{equation}
whose DOS's are displayed in Fig.\,\ref{fig:DOS_n1.0}, for different densities and several values of $t_z/t$.
For $t_z\neq0$ the DOS may be thought of two square lattice densities of states, displaced from each other by $2t_z$, as it can be seen from the positions of the van Hove singularities (VHS's). For instance, the insets of Fig.\,\ref{fig:DOS_n1.0} show the DOS as a function $t_z$, for $\langle n \rangle = 0.87$ and 0.5, respectively: a substantial increase is found in some cases. 
The fine tuning of the VHS's with a single parameter opens the possibility of increasing critical temperatures, as suggested by Eq.\,\eqref{eq:Tc_BCS}. 
Therefore, the effects of the DOS on pairing (i.e.\,whether $T_c$ increases or not) are investigated below.

%%%%%%%%%%%%%%%%  Fig 3   %%%%%%%%%%%%%%%%%%%%%%%%%%%%%%
\begin{figure}[t]
%  \centering
  \includegraphics[scale=0.35]{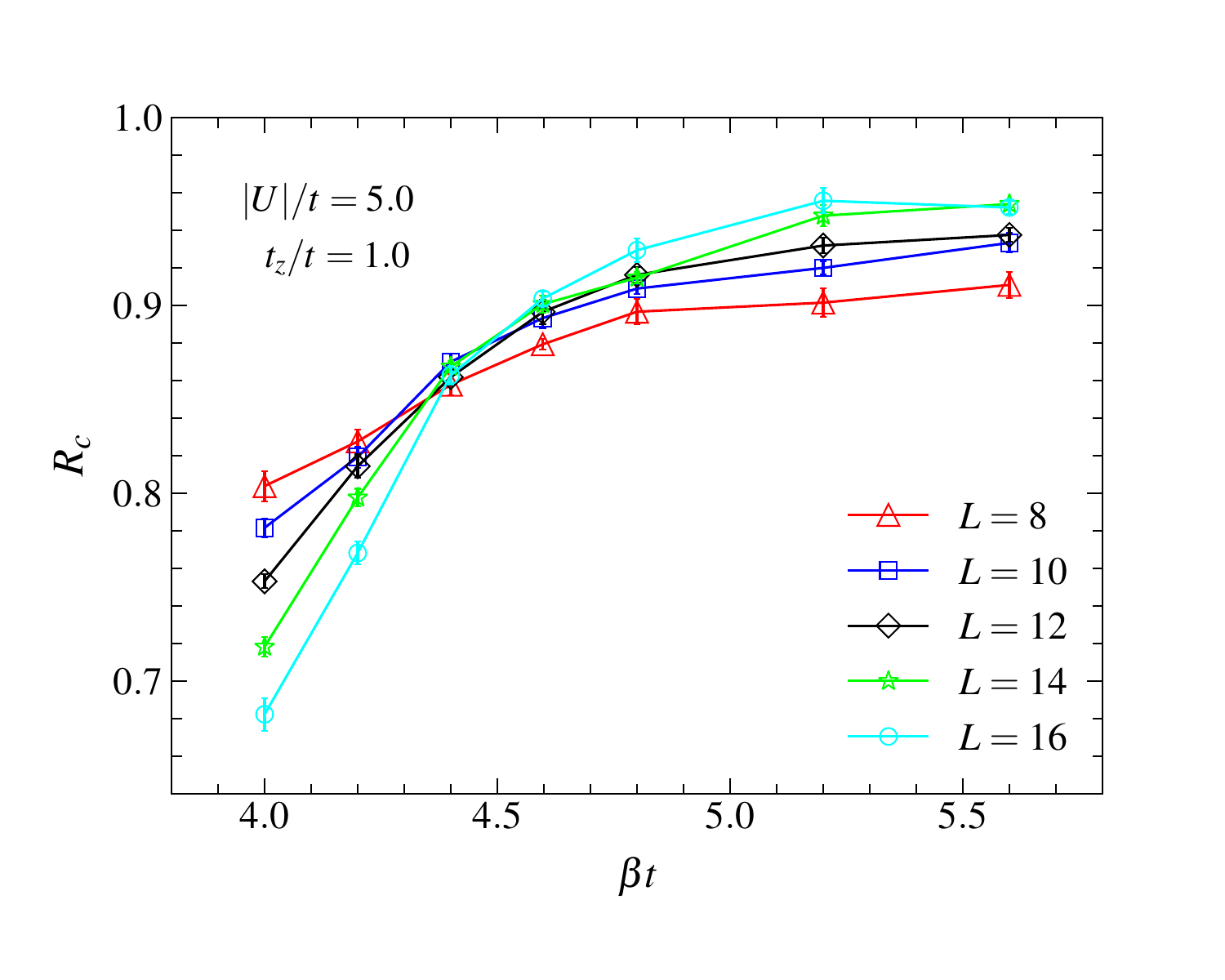}
 \caption{Correlation ratio for the AHM on a bilayer, as a function of the inverse temperature, $\beta$, for isotropic hopping and $\langle n \rangle = 0.87$. Different curves are labelled by their corresponding linear lattice sizes, $L$.}
 \label{fig:Rc_Bin0.87U5}
\end{figure}
%%%%%%%%%%%%%%%%%%%%%%%%%%%%%%%%%%%%%%%%%%%%%%%%%%%%%%%%

Recalling that the global maximum $T_c$ for the single layer lies near $\ave{n}=0.87$ \cite{Fontenele22}, we compare the effect of $t_z$ on $T_c$ both for this density and for $\ave{n}=0.5$. 
Similarly, the maximum $T_c$ over a wide range of densities occurs at $U=-5t$ \cite{Fontenele22}, so that most of our simulations  will be in the range $|U|/t=3-5$.
In this way, starting from the global maximum $T_c$ one can investigate possible paths in parameter space leading to larger slopes.

\begin{figure}[t]
  \centering
  \includegraphics[scale=0.45]{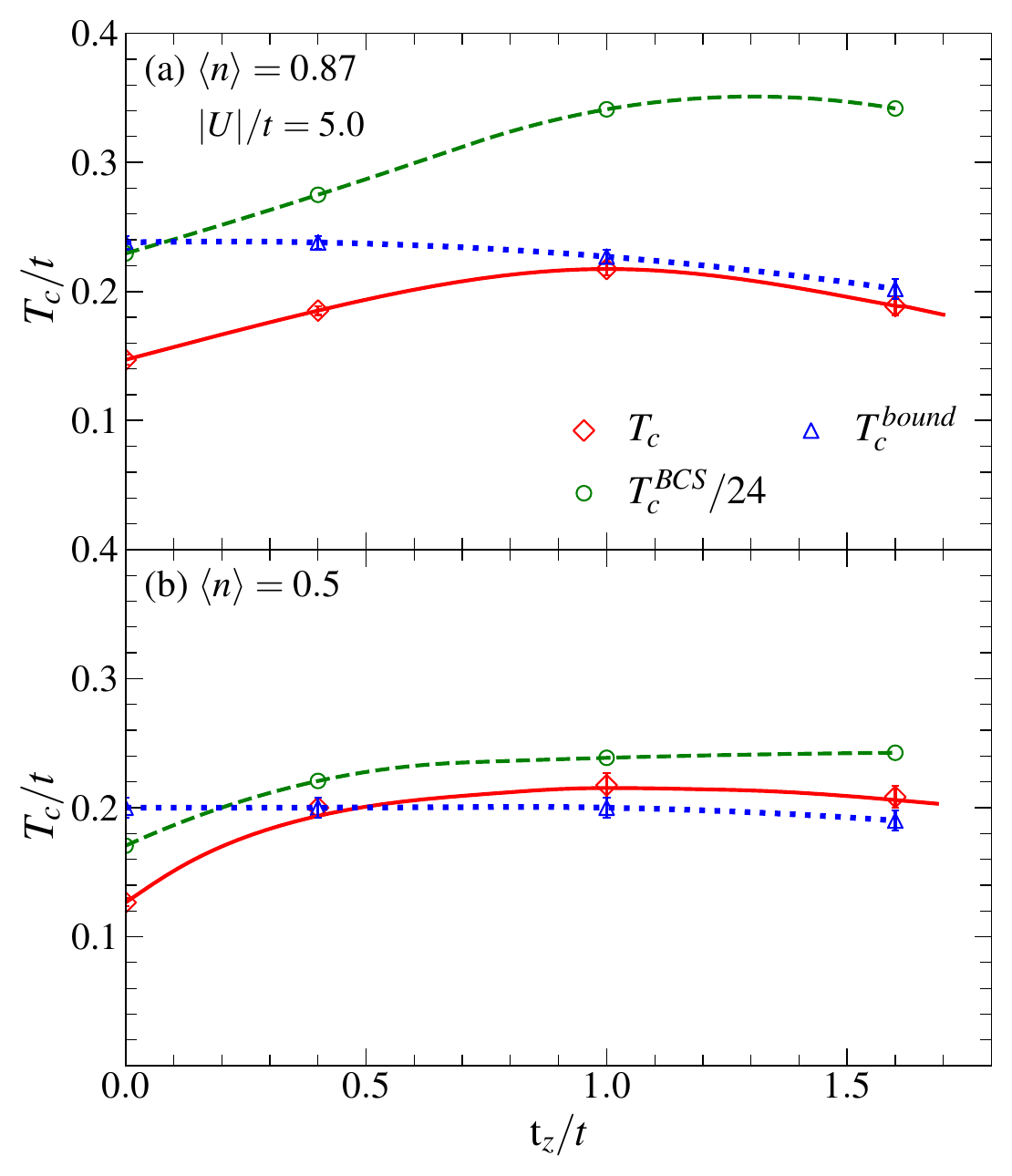}
   \caption{Critical temperature, $T_{c}/t$, for the AHM on a bilayer (red diamonds), upper bound for $T_c$ (up blue triangles) and the BCS estimate (green circles), Eq.\,\eqref{eq:Tc_BCS} as a function of interlayer hopping, $t_z/t$, for different fermionic densities, $\ave{n}$, and fixed $U$. We have arbitrarily divided $T_c^\text{BCS}$ [Eq.\,\eqref{eq:Tc_BCS}] by 24 just to lie closer to the other curves. All lines are guides to the eye. 
   }
\label{fig:Tc_Tbound_tz_n0.87n0.5_pha_diagramBi}
\end{figure}

A typical behavior of $R_c(L,\beta)$ is shown in Fig.\,\ref{fig:Rc_Bin0.87U5} for $U=-5t$, $\ave{n}=0.87$ and $t_z=t$.
We clearly distinguish two regimes of $\beta$: one in which $R_c$ decreases with increasing $L$, and another in which $R_c$ increases. 
The inversion takes place at the common crossing point, $\beta_{c}t \approx 4.5 \pm 0.1$, or $T_c/t=0.217\pm 0.005$;
for $\ave{n}=0.5$, we get $T_c/t\approx 0.217\pm 0.009$.
These values should be compared with those for the single layer, $T_c/t=0.147 \pm 0.004$ and $T_c/t = 0.126\pm 0.003$, respectively, with corresponding increase of 48\% and 72\%.
Simulations for $t_z=0.4t$ yield $T_c/t=0.185\pm0.003$ and $0.200\pm 0.008$, respectively for $\ave{n}=0.87$ and 0.5, while for $t_z=1.6t$, one gets $T_c/t=0.189\pm0.007$ and $0.208\pm 0.008$.
Figure \ref{fig:Tc_Tbound_tz_n0.87n0.5_pha_diagramBi} shows these data, and we see that the maximum $T_c$ occurs at $t_z=t$ for both fillings.
Note that we have allowed interlayer hoppings to be greater than intralayer ones, i.e.\,$t_z>t$, to broaden the search for maximum $T_c$; while in OLE this would simply demand adjustments in the relative intensity or frequency of the laser beams, in materials this could be achieved by changing the pressure from hydrostatic to uniaxial.
The decrease in $T_c$ beyond the isotropic case may be attributed to an effective dimensional reduction, since hopping along the $z$ direction becomes more favorable than hopping within the planes, thus restricting particle movement. 
This, in turn, may increase pair-scattering effects by lattice sites, thereby lowering the critical temperature. 
Note that a similar effect has been observed in a bilayer made up from a superconducting layer and a metallic layer; see, e.g.\,Ref.\,\cite{Zujev_2014}.

%%%%%%%%%%%%%%%%  Fig 4   %%%%%%%%%%%%%%%%%%%%%%%%%%%%%%
%%%%%%%%%%%%%%%%%%%%%%%%%%%%%%%%%%%%%%%%%%%%%%%%%%%%%%%%

It is also instructive to compare $T_c$ both with the BCS estimate, Eq.\,\eqref{eq:Tc_BCS}, and with a suggested upper bound~\cite{Hazra19}; see below. 
At first sight, Fig.\,\ref{fig:Tc_Tbound_tz_n0.87n0.5_pha_diagramBi} seems to indicate that $T_c^\text{BCS}$ tracks $T_c$. However, a closer look reveals that $T_c^\text{BCS}$ does not capture the position of maximum $T_c$ for either densities.
In addition, given that the (non-interacting) bandwidth varies with $t_z$, if one rescales $T_{c}^{\text{BCS}}$ by $W=8t+2t_z$, the discrepancies with respect to the location of maximum $T_c/t$ are significantly enhanced. Therefore, one cannot take  $T_c^\text{BCS}$ at face value to pinpoint actual maxima.

%%%%%%%%% Fig 5  %%%%%%%%%%%%%%%%%%%%%%%%%%%%%%%%%%%%%%%%%%%%%%%%%%
\begin{figure}[t]
  \centering
  \includegraphics[scale=0.34]{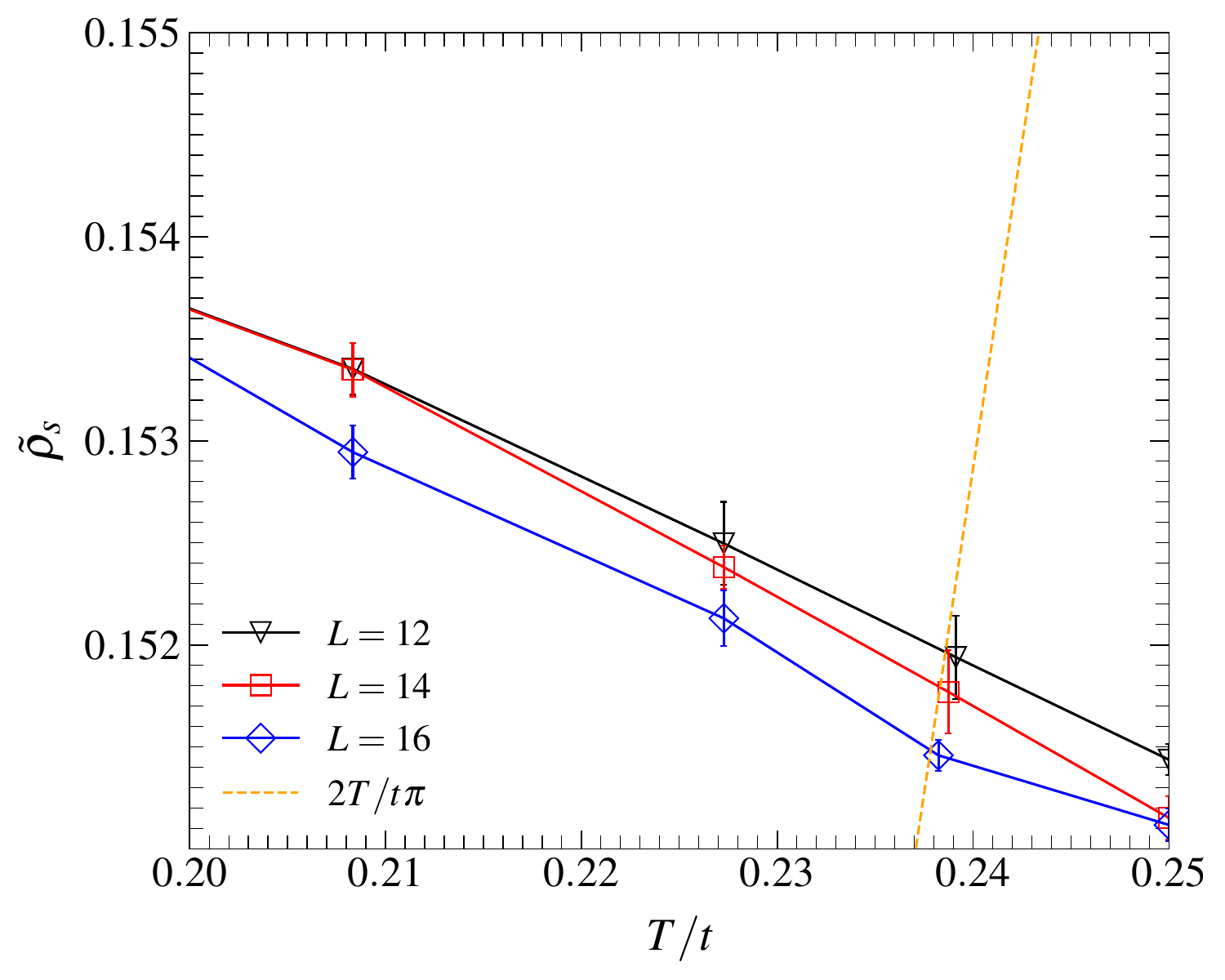}
\caption{Temperature behavior of the upper bound for the superfluid density, $\widetilde{\rho}_{s}$, at fixed $\langle n \rangle =0.87$, $U/t=-5$ and $t_{z}/t = 0.4$, and for different lattice sizes. The intercept with the (dotted) straight line $2T/t\pi$  provides an estimate for $T_c^\text{bound}$; see text.
}
 \label{fig:rhos_max}
\end{figure}

%%%%%%%%%%%%%%%%%%%%%%%%%%%%%%%%%%%%%%%%%%%%%%%%%%%%%%%%%%%%%%%%%%%%

In order to further understand the enhancement of $T_c$ presented in Fig.\,\ref{fig:Tc_Tbound_tz_n0.87n0.5_pha_diagramBi}, it is worth analyzing its upper bounds, $T_{c}^{\mbox{\tiny bound}}$, as proposed in Ref.\,\cite{Hazra19}; see also Refs.\,\cite{Hofmann22,Shi24}. We start by recalling that the superfluid density (or helicity modulus) may be obtained through~\cite{Scalapino92,Scalapino93}
\begin{align}
    \rho_{s} &= \frac{D_{s}}{4\pi e^{2}} = \frac{1}{4}\left[ \Lambda^{L} - \Lambda^{T}\right],
    \label{eq:rhos}
\end{align} 
which is proportional to the superfluid stiffness $D_s$. 
The limiting longitudinal and transverse responses are
\begin{equation}
\Lambda^L \equiv \lim_{q_x \to 0} \Lambda _{xx} (q_x, q_y=0,\omega_n=0)
\end{equation}
and
\begin{equation}
\Lambda^T \equiv \lim_{q_y \to 0} \Lambda _{xx} (q_x=0, q_y,\omega_n=0),
\end{equation}
with
\begin{align}
	\Lambda_{xx}(\mathbf{q}, \omega_n) &= \sum_{\bs{\ell}} \int_0^\beta d\tau\, e^{i\mathbf{q} \cdot \bs{\ell}} e^{i \omega_n \tau} \Lambda_{xx}(\bs{\ell},\tau),
\label{eq:lambdaq}
\end{align}
where $\omega_n=2 n \pi T$ is the Matsubara frequency, and 
\begin{align}
\label{lambda}
\Lambda_{xx}(\bs{\ell}, \tau)= \langle j_x(\bs{\ell}, \tau) j_x (0,0) \rangle,
%\Lambda_{xx}(\mathbf{l}, \tau)= \langle j_x(\mathbf{l}, \tau) j_x (0,0) \rangle,
\end{align}
where
\begin{equation}
	j_x(\bs{\ell},\tau)= e^{ {\cal H} \tau} \left[ it\sum_\sigma \left( c_{\bs{\ell} + \mathbf{\hat{x}},\sigma}^{^{\dagger}}c_{\bs{\ell},\sigma} 
		-c_{\bs{\ell},\sigma}^{^{\dagger}} c_{\bs{\ell} + \mathbf{\hat{x}},\sigma}\right)
		 \right] e ^{-{\cal H} \tau}
\end{equation}
is the $x$-component of the current density operator; see Ref.\,\cite{Scalapino92} for details.

Further, for a Berezinskii-Kosterlitz-Thouless (BKT) transition, a universal-jump relation involving the helicity modulus holds \cite{Nelson77},
\begin{align}
    T_{c} = \frac{\pi}{2} \rho_{s}^{-}
    \label{eq:Tc_rhos}
\end{align}
where $\rho_{s}^{-}$ is the value of the helicity modulus just below the critical temperature. Thus, by plotting $\rho_s$ as a function of temperature, the intercept with $2T/\pi$ provides an estimate for $T_c$, as discussed in Ref.\,\cite{Paiva04}. By the same token, a rigorous upper bound for the superfluid density (or, equivalently, for the superfluid stiffness) is given by~\cite{Hazra19} 
\begin{equation}
    \tilde{\rho}_{s} = \frac{\tilde{D}_{s}}{4\pi e^{2}} = \frac{1}{4}\Lambda^{L},
\label{eq:rhotilde}    
\end{equation}
given that $\Lambda^T$ is a positive quantity \cite{Scalapino93}. 

%%% Fig 6 %%%%%%%%%%%%%%%%%%%%%%%%%%%%%%%%%%%%%%%%%%%%%
\begin{figure}[t]
  \includegraphics[scale=0.37]{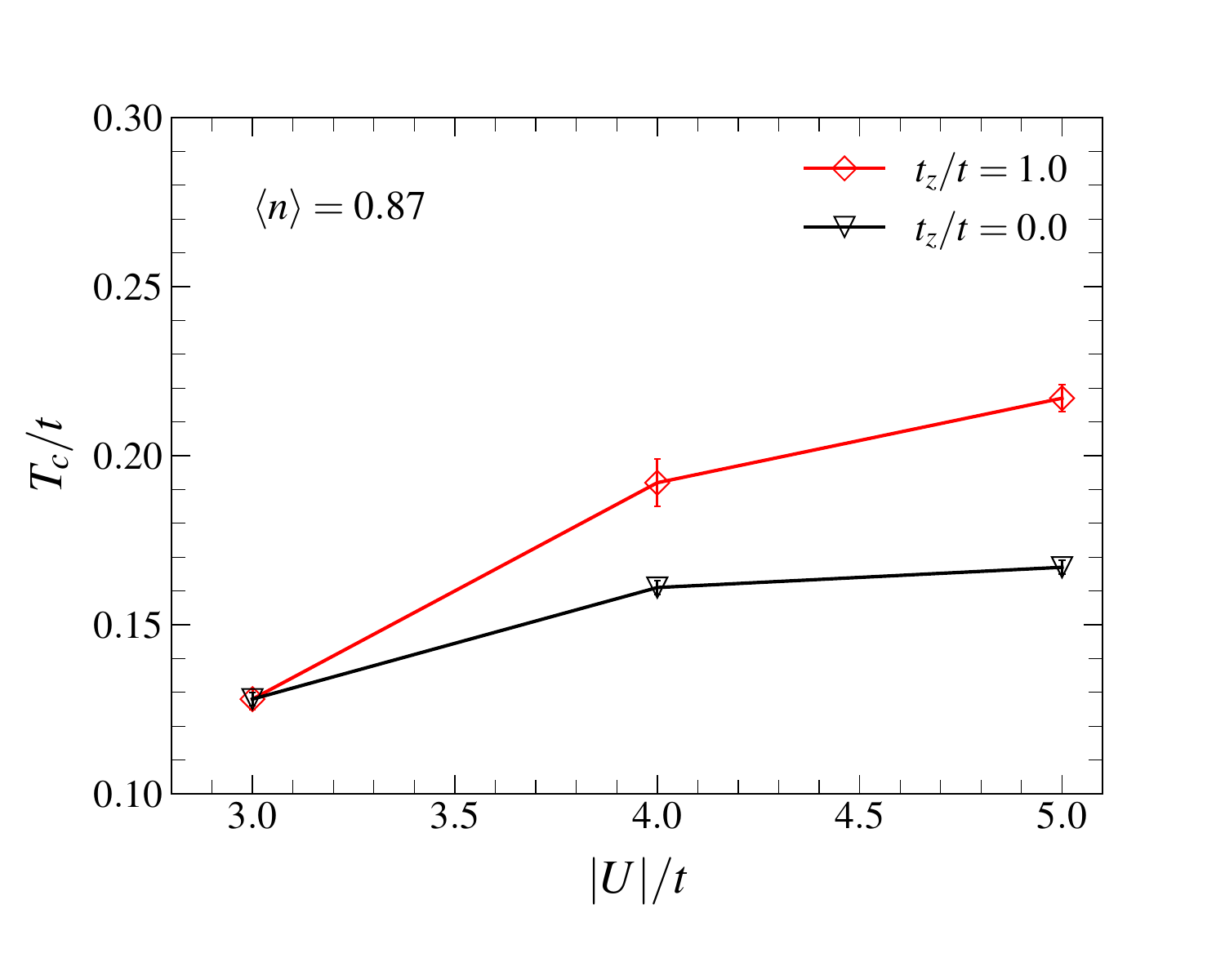}
   \caption{Critical temperature as a function of the magnitude of the attractive interaction at fixed fermionic density for both the monolayer and the isotropic (i.e., $t_{z}=t$) bilayer. 
   Lines are guides to the eye.}
\label{fig:Tc_Tbound_tzU_n0.87_pha_diagramBi}
\end{figure}
%%%%%%%%%%%%%%%%%%%%%%%%%%%%%%%%%%%%%%%%%%%%%%%%%%%%%%%%%

Since a bilayer is topologically a two-dimensional system, the above calculational strategy used for a single layer is equally applicable to the present case. 
Figure \ref{fig:rhos_max} shows typical data for $\tilde{\rho}_s(T)$ obtained from our DQMC simulations, for fixed $\ave{n}$ and $U$, and different system sizes.
Since $\tilde{\rho}_s\geq \rho_s$, and the superfluid density is a monotonically decreasing function of the temperature, at least near $T_c$ (see Ref.\,\cite{Paiva04}), the intercept of $\tilde{\rho}_s(T)$ with $2T/\pi$ will take place at some $T_c^\text{bound}\gtrsim T_c$. Figure \ref{fig:Tc_Tbound_tz_n0.87n0.5_pha_diagramBi} shows our estimates for $T_c^\text{bound}$ based on this procedure, from which we see that indeed there is not much room for $T_c$ to increase any further, in particular for $t_z \gtrsim t$. 

Let us now compare how $|U|$ influences the increase in $T_c$ in both a monolayer and an isotropic bilayer.
Figure~\ref{fig:Tc_Tbound_tzU_n0.87_pha_diagramBi} shows $T_c$ as a function of $|U|$, for a fixed $\langle n \rangle = 0.87$, where one may notice that the critical temperature increases faster for the bilayer ($t_z = 1$) than for the monolayer ($t_z = 0$).
That is, these two curves act as lower and upper bounds for $T_c(U)$ when $0< t_z/t < 1$. The observed increase in $T_{c}$ as a function of $|U|/t$ reflects the fact that the critical temperature has not yet reached its maximum value, $T_{c}^{\text{max}}$, for the specific band filling  considered. 
This behavior is consistent with results shown in Fig.\,4 of Ref.\,\cite{Fontenele22}, where for a density  $\langle n\rangle = 0.87$, $T_{c}$ increases with $|U|/t$ up to around $|U|/t=5.0$, at which point $T_{c}$ reaches a maximum, beyond which decreases steadily. It is also interesting to note that for $|U|/t=3$, $T_c$ is hardly influenced by the presence of a second layer. 

%%%% Fig. 7 %%%%%%%%%%%%%%%%%%%%%%%%%%%%%%%%%%%%%%%%%%%%%%

\begin{figure}[t]
  \centering
  \includegraphics[scale=0.34]{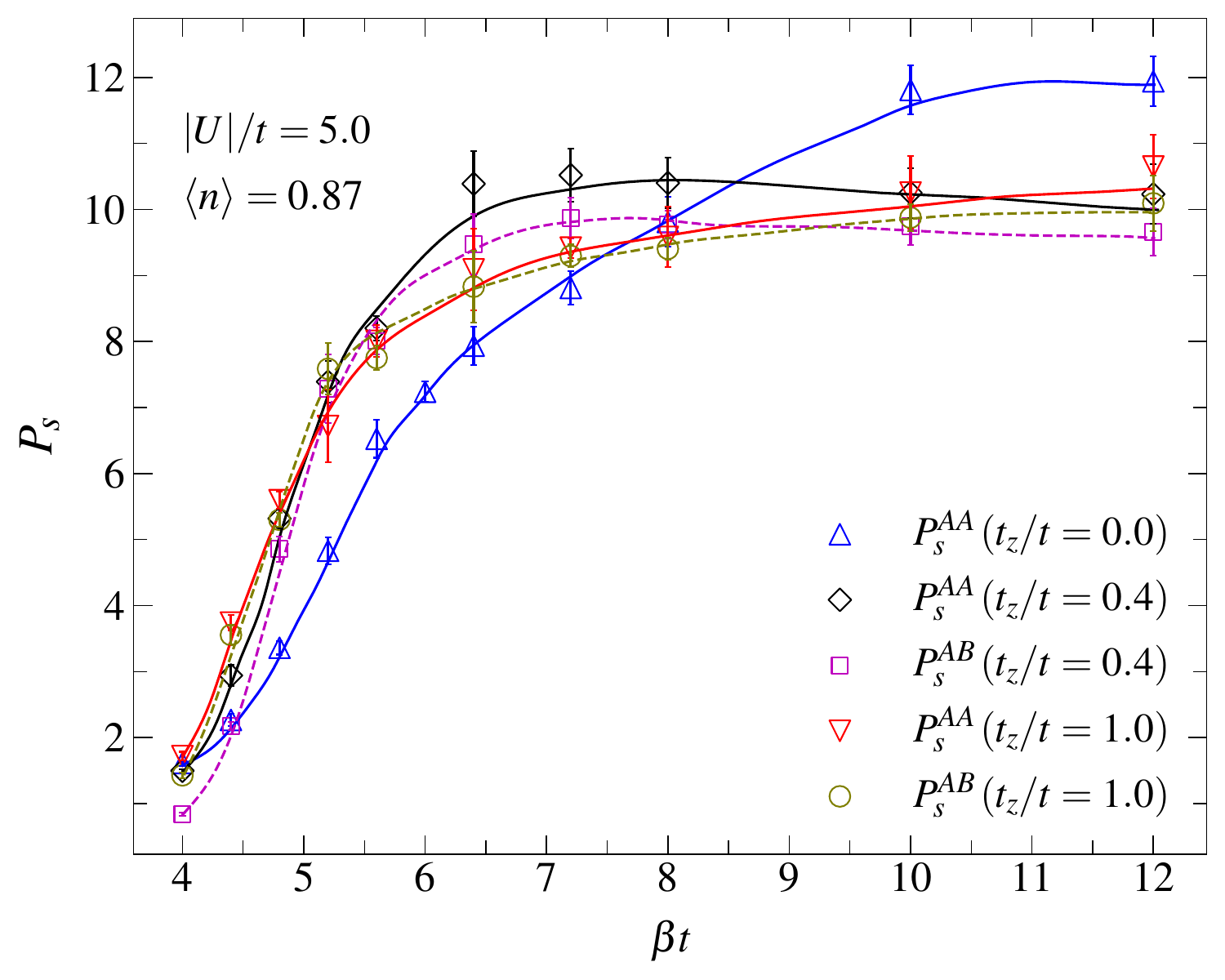}
 \caption{ 
   Layer-resolved contributions to the $s$-wave pair structure factor as functions of the inverse temperature, $\beta t$, for different values of $t_{z}$ and on an $L=16$ bilayer. 
   Lines are guides to the eye: full and dashed lines respectively refer to intralayer and  interlayer contributions. }
 \label{fig:PsAB_dif_tz_U5n0.87L16Bi}
\end{figure}

%%%%%%%%%%%%%%%%%%%%%%%%%%%%%%%%%%%%%%%%%%%%%%%%%%%%%%%%%%%%

%%%% Fig. 8 %%%%%%%%%%%%%%%%%%%%%%%%%%%%%%%%%%%%%%%%%%%%%%

\begin{figure}[h!]
  \centering
  \includegraphics[scale=0.45]{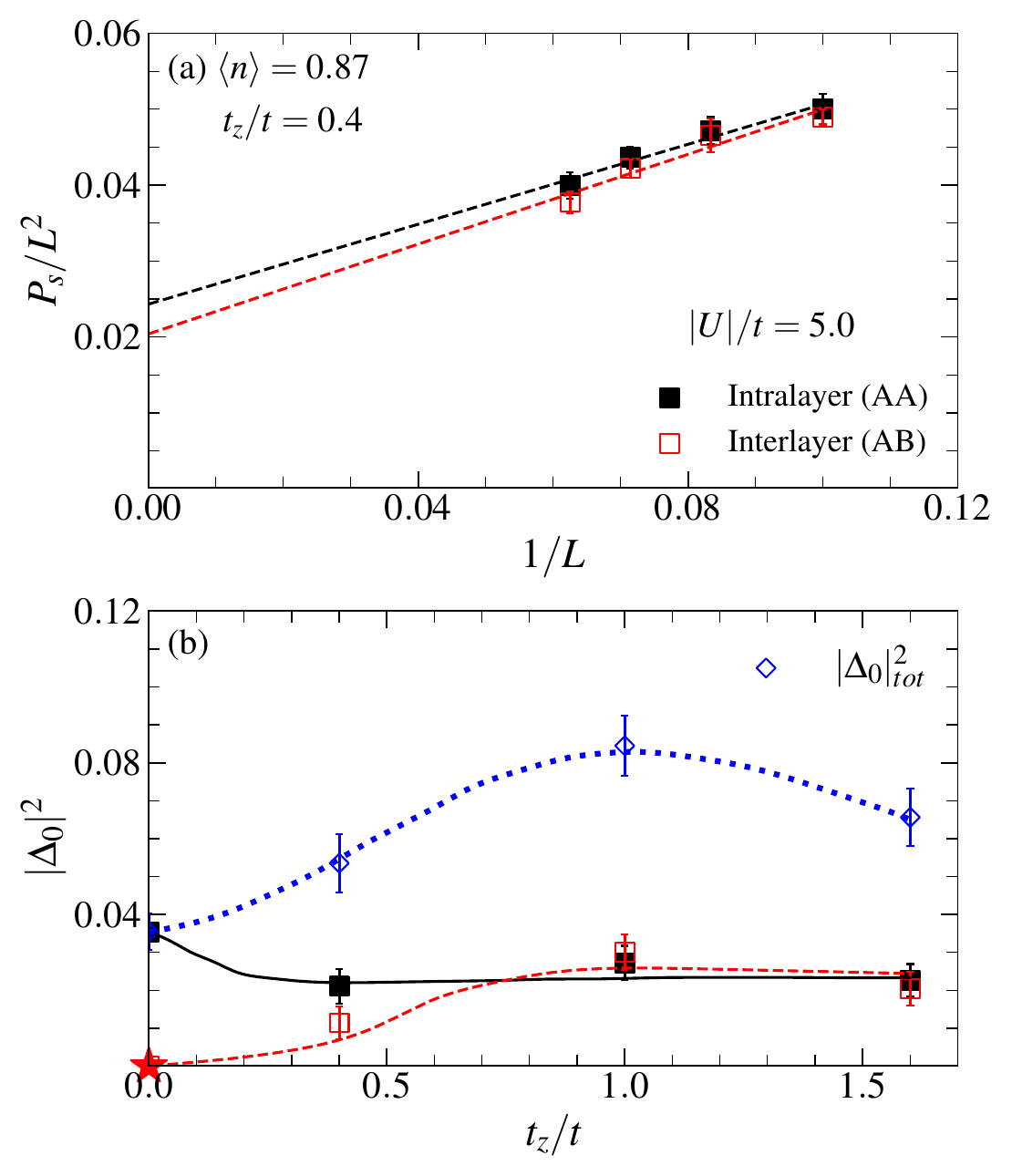}
   \caption{(a) Finite-size-scaling plot for $P_{s}$, Eq. (\ref{eq:Ps-sw}), for lattices with linear sizes $L=12, 14$ and $16$. The full and open symbols represent the intra- and inter-layer contributions to $P_{s}$. 
   (b) The intra- (full black symbols) and inter-layer (open red symbols) contributions to the squared zero temperature gap, $|\Delta_{0}|^{2}$, as functions of $t_{z}/t$. In addition, the total $|\Delta_{0}|^{2}$, as a function of $t_{z}/t$ is show (blue diamonds).}
 \label{fig:Ps_FSS_U5n087}
\end{figure}

%%%%%%%%%%%%%%%%%%%%%%%%%%%%%%%%%%%%%%%%%%%%%%%%%%%%%%%%%%%%

At this point, one may wonder which is the division of labor between intra- and interlayer superconducting correlations.
In order to shed light into this issue, we restrict the sums in the pair structure factor, Eq.\,\eqref{eq:Ps-def}, to sites within the same layer, call it $P_s^\text{AA}$, and to sites joining the two layers along the $z$ direction, $P_s^\text{AB}$, in such way that
\begin{equation}
    P_{s}=2P_s^{\text{AA}}+P_s^{\text{AB}}.
\end{equation}
Typical data for $P_s^\text{AA}$ and $P_s^\text{AB}$ as functions of the inverse temperature are shown in Fig.\,\ref{fig:PsAB_dif_tz_U5n0.87L16Bi}.
By extracting similar data for other system sizes, we can plug their limiting ($\beta\to\infty$) values into the Huse scaling \cite{Huse88},
\begin{equation}
      \frac{P_s}{L^2}= |\Delta_0|^2 + \frac{B}{L}, \ \ \ L\gg 1,\ T\to 0,
\label{eq:Ps-sw}
\end{equation}
to obtain the separate contributions to the superconducting gap at zero temperature \cite{Moreo91,Paiva04}; $B$ is a non-universal constant. 

Figure \ref{fig:Ps_FSS_U5n087}(a) illustrates the extrapolation to $L\to\infty$ in a scaling plot. 
Interestingly, when we plot the separate gaps as functions of $t_z$ in Fig.\,\ref{fig:Ps_FSS_U5n087}(b) for fixed $U$ and $\ave{n}$, we see that the intralayer contribution initially decreases before stabilizing in a value smaller than that for a monolayer. 
By contrast, the interlayer contribution rises from zero and stabilizes at roughly the same value as the one for the monolayer. 
Equation \eqref{eq:Ps-sw} allows us to add the contributions to the total $|\Delta_0|^2$, and  Fig.\,\ref{fig:Ps_FSS_U5n087}(b) shows that the latter displays a maximum near the isotropic limit, $t_z=t$, similarly to $T_c$.
Thus, the key point is that the correlations between layers lead to a stronger order parameter than in a monolayer, which, in turn, results in higher critical temperatures.

%%%%%%%%%%%%%%%%%%%%%%%%%%%%%%%%%%%%%%%%%%%%%%%%%%%%%%%%%%%%%
\subsection{Pairing temperature scale}
\label{ssec:bilayer-Tp}

%% fig 9 %%%%%%%%%%%%%%%%%%%%%%%%%%%%%%%%%%%%%%%%%%
\begin{figure}[t]
  \centering
  \includegraphics[scale=0.4]{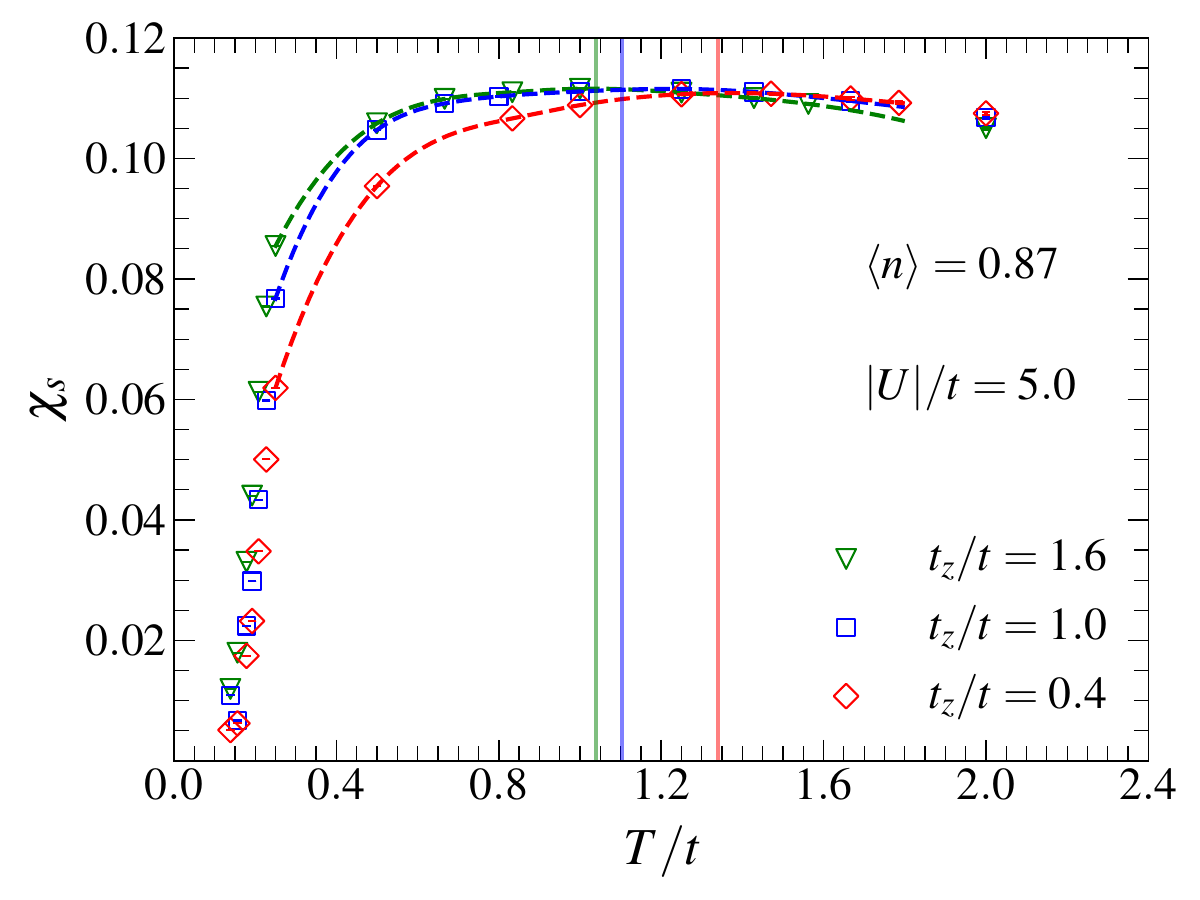}
    \caption{The uniform spin susceptibility for the bilayer, as a function of temperature for  $\langle n\rangle = 0.87$, $|U|/t = 5.0$, and different values of the interplane hopping $t_{z}$; the  linear lattice size is $L=16$. 
    Dashed lines are guides to the eye, and the vertical lines indicate the temperatures at which the downturn in $\chi_s$ occurs.}
 \label{fig:Chi_s_tz_n087U5}
\end{figure}
%%%%%%%%%%%%%%%%%%%%%%%%%%%%%%%%%%%%%%%%%%%%%%%%%%%%%%

%%%% Fig 10 %%%%%%%%%%%%%%%%%%%%%%%%%%%%%%%%%%%%%%%%%%%

\begin{figure}[h!]
  \centering
  \includegraphics[scale=0.45]{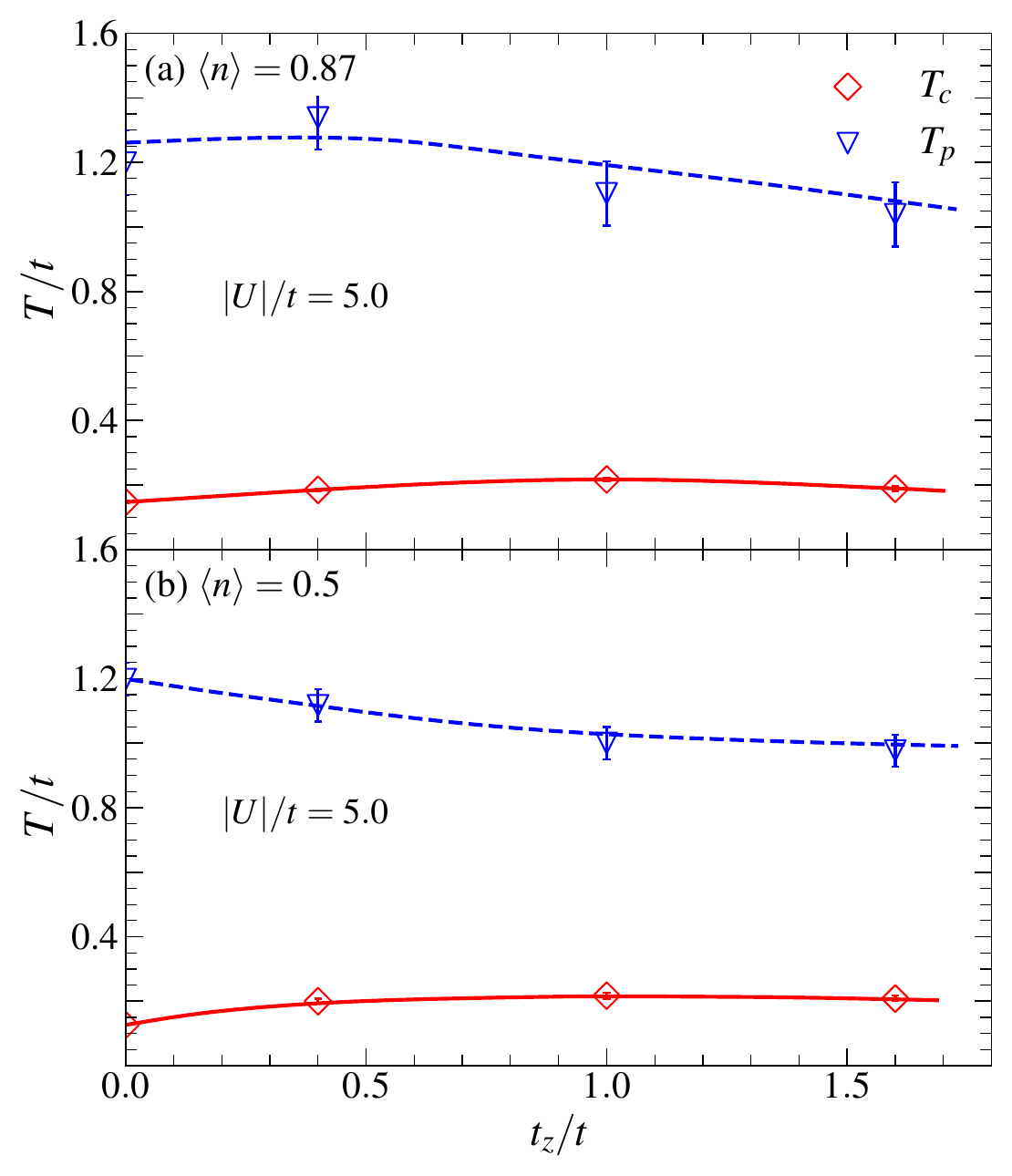}
    \caption{Critical ($T_c$) and pairing ($T_p$) temperatures (in units of $t$) as functions of $t_{z}/t$, obtained from our DQMC simulations for a bilayer with linear size $L = 16$, and 
    (a) $\langle n\rangle = 0.87$; (b)$\langle n\rangle = 0.5$. 
    Lines are guides to the eye.  }
 \label{fig:Tc_Tp_tz_bilayer}
\end{figure}
%%%%%%%%%%%%%%%%%%%%%%%%%%%%%%%%%%%%%%%%%%%%%%%%%%%%%%%

%%% Fig 11 %%%%%%%%%%%%%%%%%%%%%%%%%%%%%%%%%%%%%%%%%%%%%%%%%%%%%%%%%%%%%%%%%%
\begin{figure*}[t]
  \centering
  \includegraphics[scale=0.575]{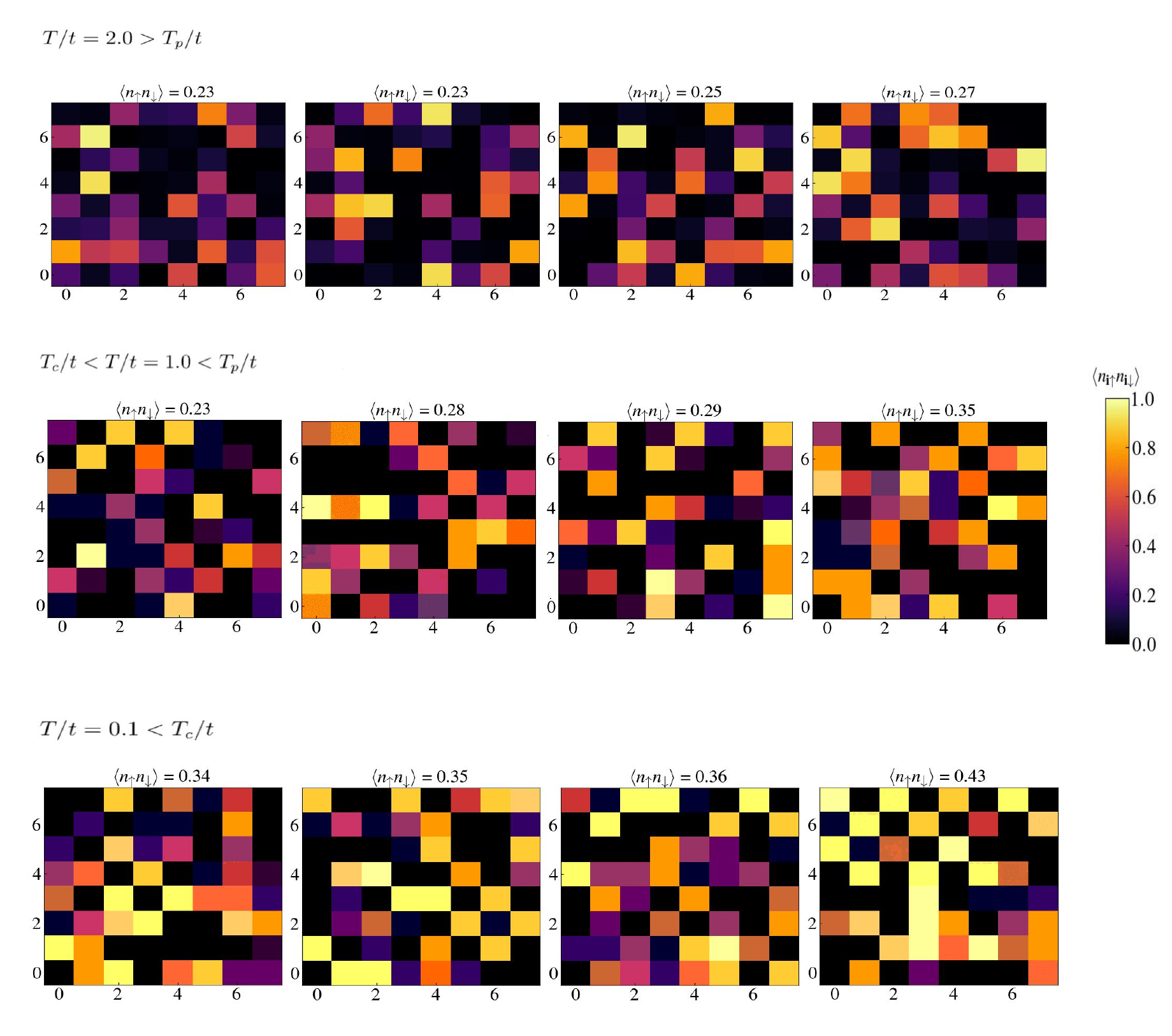}
 \caption{Snapshots of the double occupancy, $d_\iv$, throughout one of the layers. 
    Labels of horizontal and vertical axes denote site coordinates on an $L=8$ layer, for $\langle n\rangle = 0.87$, $U/t = -5$ and $t_{z}=t$.
    The rows show typical distributions at different temperatures.
    The calculated double occupancy for each realization appears on top of each panel, and the color map on the right hand side shows the scale for $d_\iv$. }
 \label{fig:snapstz10}
\end{figure*}

%%%%%%%%%%%%%%%%%%%%%%%%%%%%%%%%%%%%%%%%%%%%%%%%%%%%%%%%%%%%%%%%%

%%% Fig 12 %%%%%%%%%%%%%%%%%%%%%%%%%%%%%%%%%%%%%%%%%%
\begin{figure}[t]
  \centering
\includegraphics[scale=0.31]{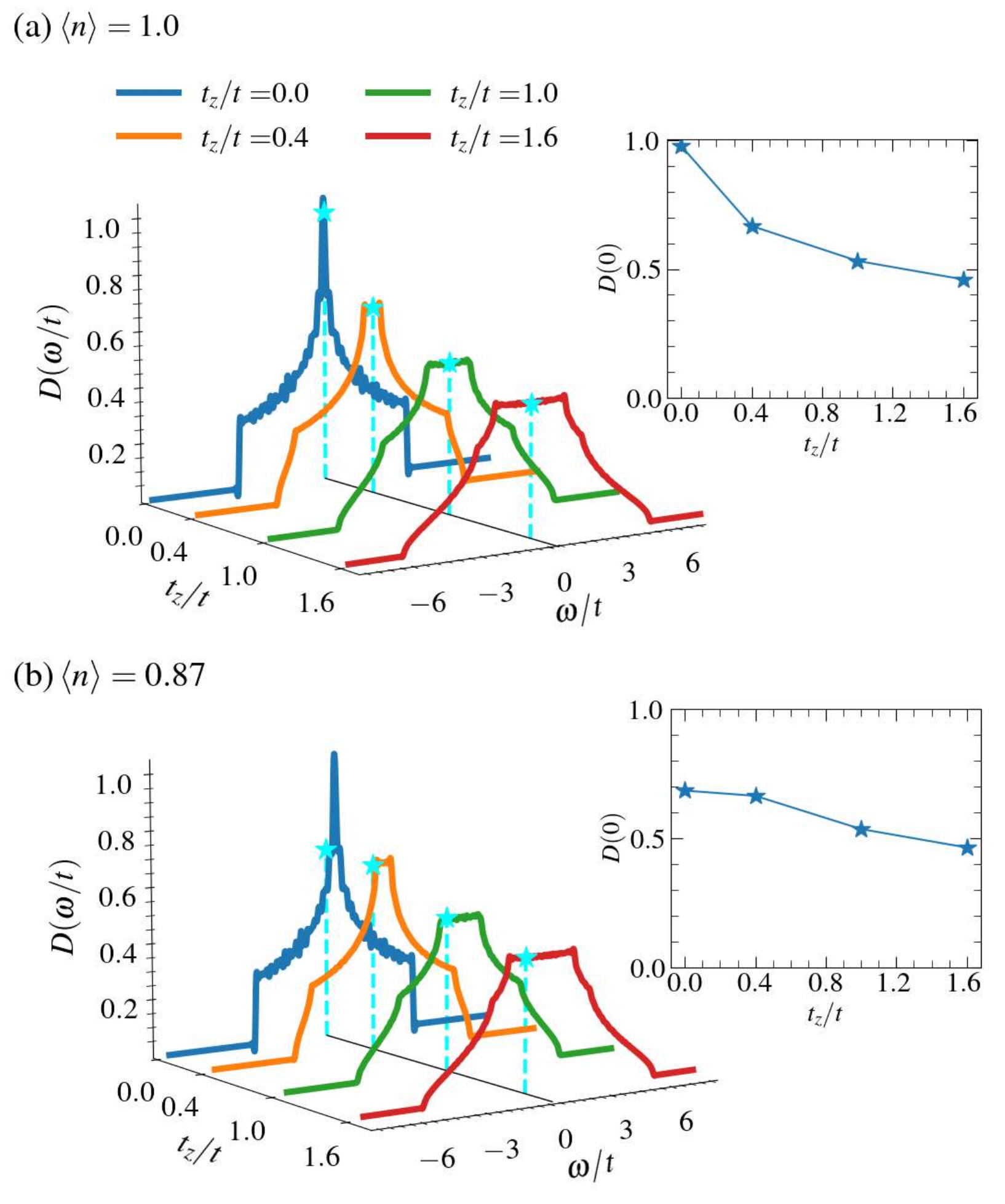}
    \caption{Same as Fig.\,\ref{fig:DOS_n1.0}, but for a simple cubic lattice:
    (a) $\langle n \rangle = 1.0$ and (b) $\langle n \rangle = 0.87$.    }
 \label{fig:DOS_cubic}
\end{figure}

%%% Fig 13 %%%%%%%%%%%%%%%%%%%%%%%%%%%%%%%%%%%%%%%%%%%%%
\begin{figure}[t]
  \includegraphics[scale=0.33]{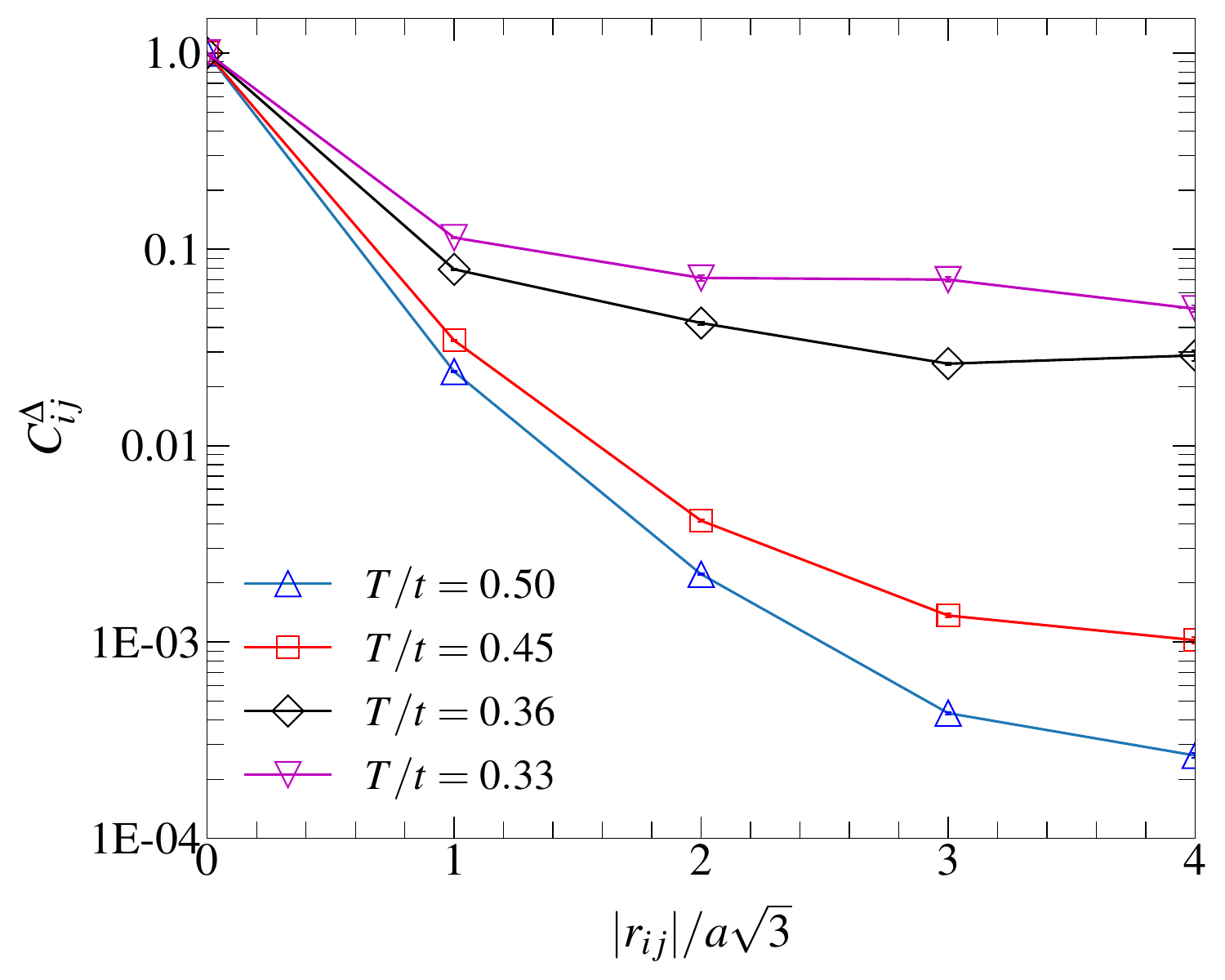}
   \caption{ 
   Log-linear plot of the pairing correlation function \textit{vs.} distance along the diagonal of an $8 \times 8 \times 8$ cubic lattice, for different inverse temperatures, $\beta t$, for a half-filled band, isotropic hopping $t_{z}/t = 1$, and $U/t = -8.0$. 
    Due to periodic boundary conditions, the farthest distance is $4a\sqrt{3}$, with $a$ being the lattice spacing.
    }
 \label{fig:PsR_n1.0tz1}
\end{figure}

%%%% Fig 14 %%%%%%%%%%%%%%%%%%%%%%%%%%%%%%%%%%%%%%%%%%%%%%%%%
\begin{figure}[t]
  \includegraphics[scale=0.35]{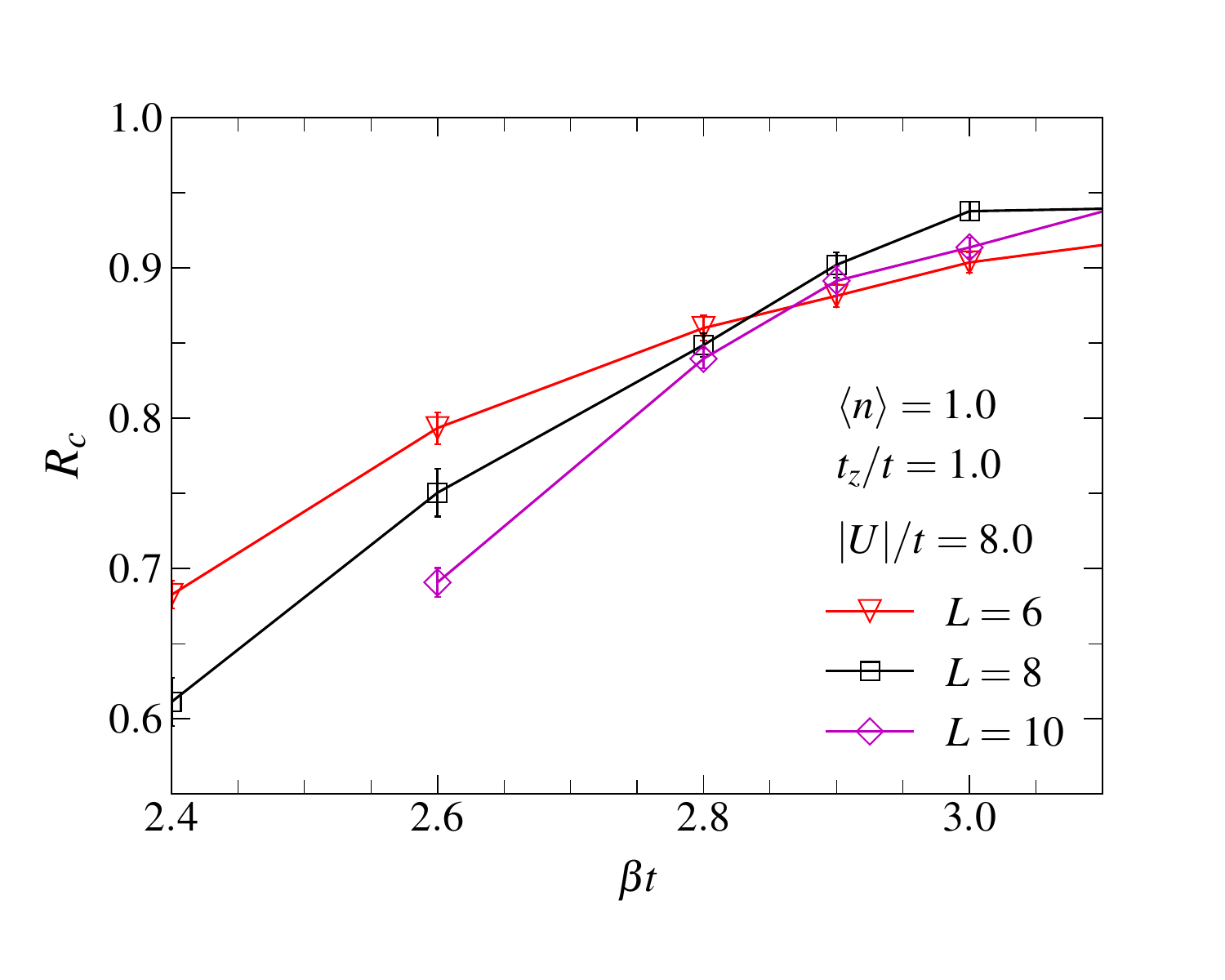}
   \caption{Correlation ratio as a function of the inverse temperature, $\beta t$, for different linear cubic lattice sizes. The inverse critical temperature is estimated as $\beta_{c}t = 2.85 \pm 0.05$.}
 \label{fig:Rc_n1.0}
\end{figure}

%%%%%%%% Fig 15 %%%%%%%%%%%%%%%%%%%%%%%%%%%%%%%%%%%%%%
\begin{figure}[t]
  \includegraphics[scale=0.45]{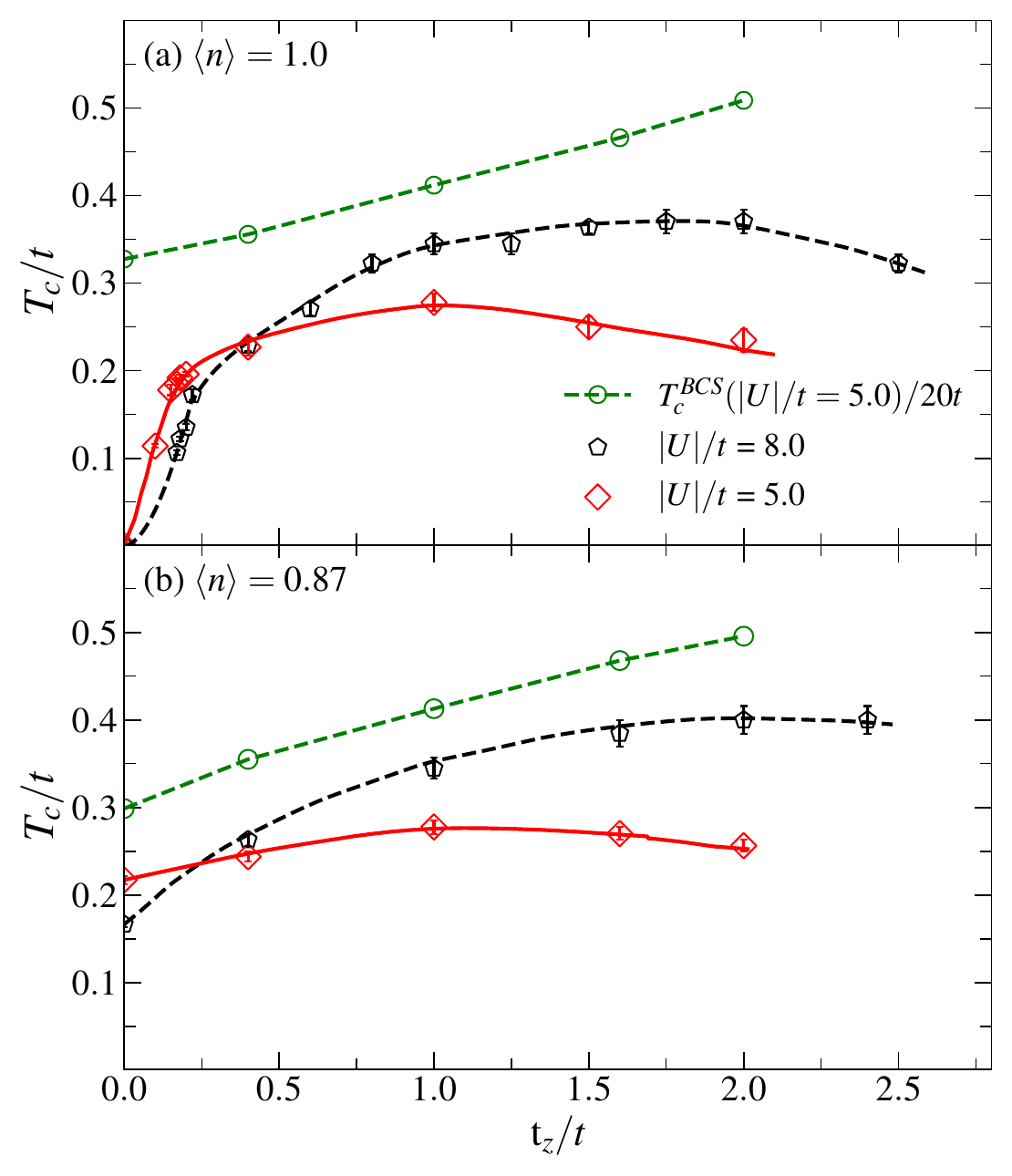}
    \caption{DQMC estimates for the 3D critical temperature, $T_c$, as a function of $t_{z}/t$: (red) diamonds and (black) pentagons correspond to $U=-5t$ and $U=-8t$, respectively; scaled data for $T_c^\text{BCS}$, Eq.\,\eqref{eq:Tc_BCS}, for $U=-5t$, are shown as (green) circles. We have arbitrarily divided $T_c^\text{BCS}$ by 20 just to lie closer to the other curves. The panels correspond to the  densities shown. All lines are guides to the eye.}
\label{fig:Tc_tz_n1.0_pha_diagram}
\end{figure}

%%%%%%%%%%%%%%%%%%%%%%%%%%%%%%%%%%%%%%%%%%%%%%%%%%%%%%%

%%%% Fig 16 %%%%%%%%%%%%%%%%%%%%%%%%%%%%%%%%

\begin{figure}[t]
  \centering
  \includegraphics[scale=0.45]{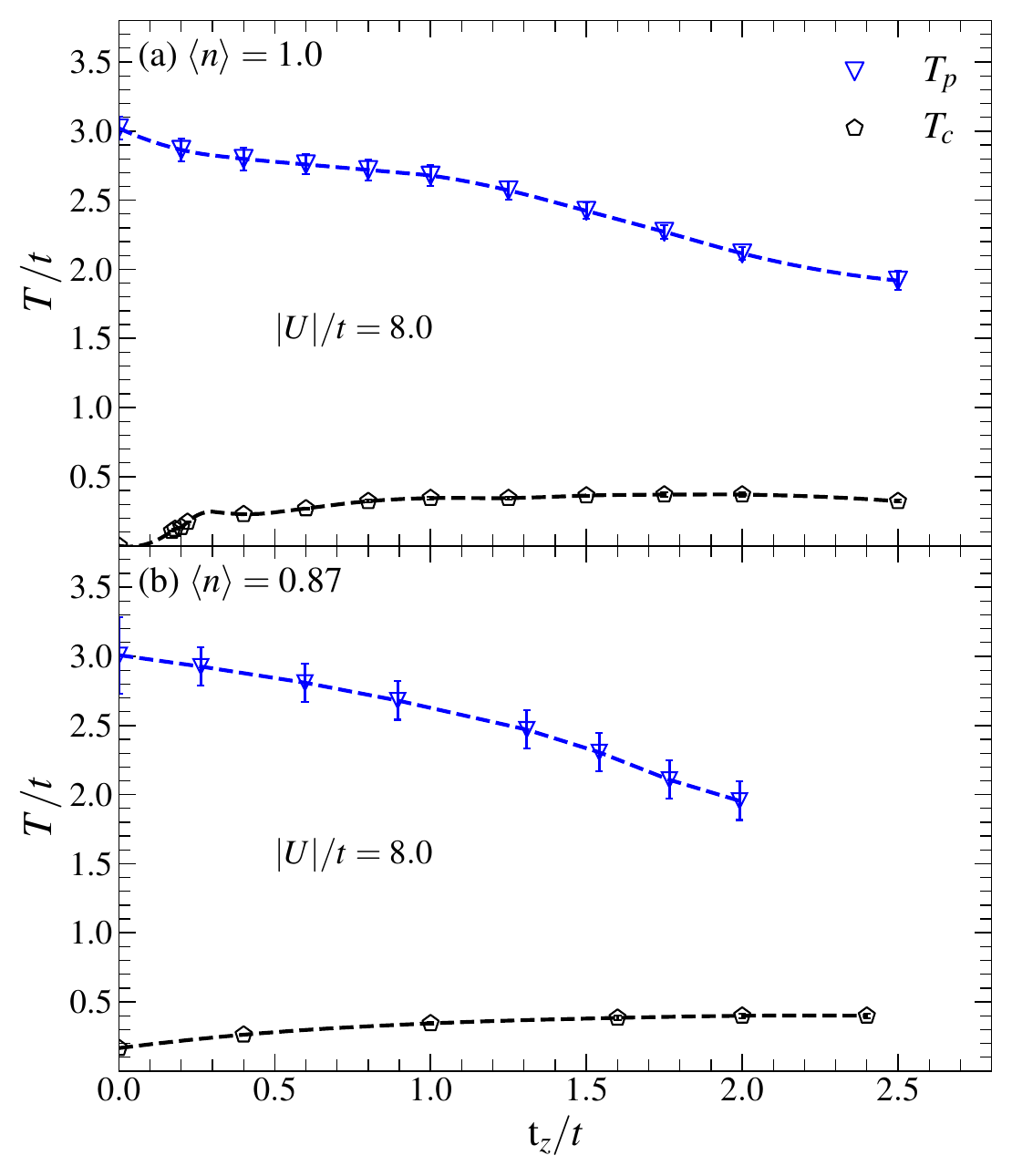}
    \caption{Critical, $T_c$, and pairing, $T_p$, temperatures (in units of $t$) as functions of  $t_{z}/t$, obtained from our DQMC simulations for a cubic lattice with linear lattice size $L = 8$, at (a) half filling and (b) for $\langle n\rangle = 0.87$, for $|U|/t = 8.0$. 
    Dashed lines are guides to the eye.
    }
 \label{fig:Tc_Tp_TcBCS_tz_U8}
\end{figure}

%%%%%%%%%%%%%%%%%%%%%%%%%%%%%%%%%%%%%%%%%%%%%%%

As mentioned in the Introduction, one of the interesting features of the AHM is the emergence of a pairing temperature scale above $T_c$ which signals the formation of pairs, but not their coherence. 
Our purpose here is to investigate how the addition of another layer affects the pairing temperature.
One manifestation of pair formation is through a gap opening for spin excitations, hence as a downturn in the uniform magnetic susceptibility $\chi_{s}$ as the temperature is lowered \cite{Micnas90,Randeria92,dosSantos94,Wlazlowski13,Tajima14,Fontenele22}.

Figure \ref{fig:Chi_s_tz_n087U5} shows our typical DQMC data for the temperature dependence of the uniform susceptibility, $\chi_s$ [see Eq.\,\eqref{eq:magsusc}], for a bilayer with $|U|/t = 5.0$ and $\langle n\rangle = 0.87$, for different interplane hoppings; the linear lattice size is kept fixed at $L=16$.  
We note that, for fixed $U$, $\chi_{s}$ drops steadily below some temperature, whose location depends on $t_{z}$. 
Following the idea that this drop signals the formation of local pairs within some temperature scale~\cite{Randeria92,dosSantos94,Wlazlowski13,Tajima14}, we adopt the position of the maximum in $\chi_s$ as the pairing scale, $T_p$. 
As Fig.\,\ref{fig:Chi_s_tz_n087U5} shows, for strong couplings the maxima can be quite broad, so that their position is determined by inspection of the actual numerical output for $\chi_s$, taking into account its error bars; this yields a range of temperatures within which the maximum lies. 
This somewhat flexible definition is in line with the fact that we are dealing with a crossover, hence a temperature scale, not with a sharp transition.

Estimates for $T_p$ as a function of $t_z$ are shown in Fig.\,\ref{fig:Tc_Tp_tz_bilayer} for two different densities; for comparison, data for $T_c(t_z)$ are also shown.
We see that the overall tendency of $T_p$ is to decrease very slowly with $t_z$, while, as mentioned before, $T_c$ displays a broad maximum around $t_z=t$. 
From this we may conclude that, to some extent,   interlayer hopping favors pair breaking if they are not part of a condensate, but suppresses pair breaking if in a condensate. 

We have also studied the distribution of doubly occupied sites, through 
\begin{equation}
d_\iv\equiv\ave{n_{\iv\uparrow}n_{\iv\downarrow}},
    %\langle n_{\mathbf{i},\uparrow}n_{\mathbf{i},\downarrow}\rangle = \langle c_{\mathbf{i},\uparrow}^{\dagger}c_{\mathbf{i},\uparrow}c_{\mathbf{i},\downarrow}^{\dagger}c_{\mathbf{i},\downarrow} \rangle~.
    \label{eq:douboc}
\end{equation}
and its average over the lattice sites of the bilayer,
\begin{equation}
\ave{d}\equiv\ave{n_\uparrow n_\downarrow}
    \equiv\frac{1}{N_s}\sum_\iv d_\iv,
\end{equation}
with $N_s=2L^2$.
Figure \ref{fig:snapstz10} displays snapshots of $d_\iv$ for $\ave{n} = 0.87$ and $|U|/t = 5.0$ for an $L=8$  bilayer with $t_z=t$ within three ranges of temperature, namely the metallic phase, $T>T_p$, the pseudogap regime, $T_c < T < T_p$, and the superconducting phase, $T<T_c$.  
As expected, we see that $\ave{d}$  typically increases as the temperature is lowered, with the boost being more pronounced as the pairs condense into the superconducting phase.
We also see that there is no significant change in the range of values of $\ave{d}$.
By contrast, the distribution of doublons seems to follow a more pronounced checkerboard pattern in the latter case.
In the superconducting phase the checkerboard pattern is even more pronounced than in the pseudogap phase, in the sense that one finds sites with large $\ave{d_\iv}$ surrounded by sites with small $\ave{d_\iv}$.
One may attribute this to the presence of more tightly bound pairs. 

\section{The Simple Cubic lattice}
\label{sec:cubic}

Having established the quantitative influence of hopping to another layer on the critical temperature, it is instructive to compare with the behavior of the AHM on a simple cubic lattice, but with a variable hopping along one of the directions, say the $z$ direction, as in the bilayer.
Recalling that now $T_c\neq0$ for $t_z\neq0$  at half filling, we will discuss this case, in addition to $\ave{n}=0.87$, which is where the maximum $T_c$ occurs for the two-dimensional case.   

Starting with the non-interacting case, Fig.\,\ref{fig:DOS_cubic} shows the density of states for different values of $t_z/t$, and for the two fillings mentioned above.
We see how the van Hove singularity at the two-dimensional particle-hole symmetry point evolves to the typical three-dimensional DOS. 
As a consequence, the DOS at the corresponding Fermi energies monotonically decrease with $t_z/t$; see the insets in each case.

Figure \ref{fig:PsR_n1.0tz1} shows the decay of pairing correlations with the distance for the cubic lattice at half filling.
It is interesting to note that the length scale of decay increases by two orders of magnitude as the temperature is lowered within the interval shown, signalling the onset of superconducting long-range order.   
In order to estimate $T_c$ through our QMC simulations, we adopt a procedure similar to that for the bilayer, namely the crossing of the correlation ratio, Eq.\,\eqref{eq:Rc_def}.
Typical results are shown in Fig.\,\ref{fig:Rc_n1.0}, and by drawing similar plots for other values of $U$, $t_z$, and $\ave{n}$ one determines $T_c(t_z/t)$ as displayed in Fig.\,\ref{fig:Tc_tz_n1.0_pha_diagram}. 
We see that for $U=-5t$ the maximum $T_c$ occurs at $t_z=t$ for both fillings; therefore, in this case varying $t_z$ away from the isotropic limit does not lead to higher $T_c$'s.
By contrast, for $U=-8t$ the maximum $T_c$ occurs at larger values of $t_z$ as the filling is decreased: there is an increase of $T_c / t = 0.34 \pm 0.01$, for $t_z = t$, from $T_c / t = 0.40 \pm 0.02$ for $t_z = 2t$, which is about 16\% over the value for the isotropic case.
This may be attributed to the fact that for stronger couplings the pairs are more strongly bound, hence more resistant to the suppressing effects of channelling along the $z$-direction.

Figure \ref{fig:Tc_tz_n1.0_pha_diagram} shows that the maximum increase in $T_c$ as $|U|$ goes from $5t$ to $8t$ is of about $50\%$, reachable when $t_z>1.5 t$. In addition, Fig.\,\ref{fig:Tc_tz_n1.0_pha_diagram} also shows an interesting feature: a crossing of the $T_c(t_z/t)$ curves for small values of $t_z$. In order to understand this, let us focus on $\ave{n}=0.87$. For the square lattice the maximum $T_c$ occurs for $U=-5t$ \cite{Fontenele22}, namely $T_c\approx 0.22t$, while for $U=-8t$, $T_c\approx 0.16t$; see data for $t_z=0$ in  Fig.\,\ref{fig:Tc_tz_n1.0_pha_diagram}(b). On the other hand, for the isotropic simple cubic lattice the maximum $T_c$ occurs for $|U|/t \gtrsim 8$, as it can be inferred from the data for $\ave{n}=0.8$ in Ref.\,\cite{dosSantos94}. 
The crossing is therefore a consequence of the dimensional crossover driven by $t_z$. 
In other words, the crossing at  $t_z \approx 0.25t$ for $\ave{n}=0.87$ seems to separate two regimes: one in which interplane pairing correlations are significantly less relevant than the intraplane ones.
Similar arguments should also account for the crossing at half filling, despite the fact that $T_c$ is suppressed to zero for $t_z=0$. 

Similarly to what we did for the two-dimensional geometries, we now probe the effectiveness of the BCS estimate for the anisotropic simple cubic lattice. 
Figure \ref{fig:Tc_tz_n1.0_pha_diagram} compares the dependence of $T_c/t$ and $T_c^\text{BCS}/W$ with $t_z/t$, for fixed $U=-5t$, and for two densities; $W = 8t + 4t_z$ is the bandwidth. 
For half filling, $T_c^\text{BCS}$ does not satisfy the Mermin-Wagner theorem for $t_z=0$, as expected. In addition, one sees that $T_c^\text{BCS}$ does not track our DQMC estimates for $T_c$ for both fillings.

Let us now discuss how the pairing temperature is affected by a variable hopping along the $z$-direction.
Our DQMC data are shown in Fig.\,\ref{fig:Tc_Tp_TcBCS_tz_U8} for fixed $U=-8t$, and for different fillings; $T_c$ is also plotted for comparison.
We see that $T_p$ decreases much more steadily with $t_z$ than for the bilayer; this leads one to conclude that confinement along the $z$-direction favors pair formation, but not necessarily pair condensation. 
The data also show that the behavior of $T_p$ with $t_z$ is not too dependent on $\ave{n}$ within the range considered.  

\begin{figure}[t]
  \centering
  \includegraphics[scale=0.5]{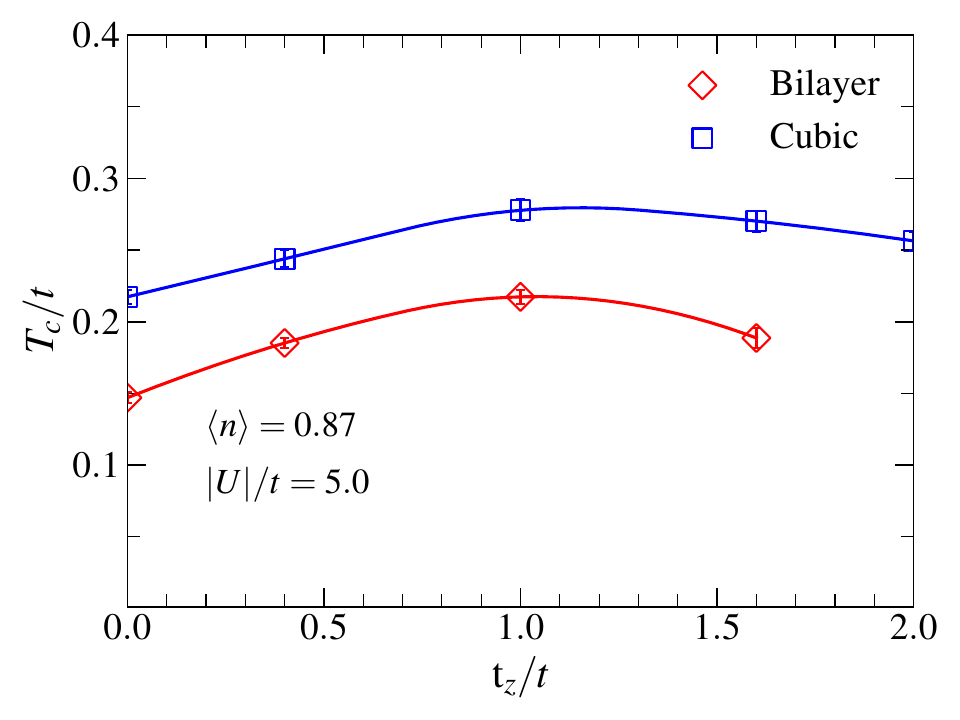}
    \caption{Comparison of our DQMC data for $T_c$ on the bilayer and on the simple cubic lattice.
    }
 \label{fig:Tc_Tp-2D3D}
\end{figure}

And, finally, Fig.\,\ref{fig:Tc_Tp-2D3D} compares the effect of anisotropy on $T_c$ for both the bilayer and the simple cubic lattice. 
We see that away from half filling $Tc$ is about 30\% higher for the 3D lattice.

%%%% Fig 17 %%%%%%%%%%%%%%%%%%%%%%%%%%%%%%%%

%%%%%%%%%%%%%%%%%%%%%%%%%%%%%%%%%%%%%%%%%%%%%%%

%%%%%%%%%%%%%%%%%%%%%%%%%%%%%%%%%%%%%%%%%%%%%%%%%%%%%%%%%%%%%%%%%%
%%%%%%%%%%%%%%%%%%%%%%%%%%%%%%%%%%%%%%%%%%%%%%%%%%%%%%%%%%%%%%%%%%
\section{Conclusions}
\label{sec:conc}
%%%%%%%%%%%%%%%%%%%%%%%%%%%%%%%%%%%%%%%%%%%%%%%%%%%%%%%%%%%%%%%%%%
%%%%%%%%%%%%%%%%%%%%%%%%%%%%%%%%%%%%%%%%%%%%%%%%%%%%%%%%%%%%%%%%%%

We have examined layering as a possible route to increase the superconducting critical temperature in the attractive Hubbard model. 

In two dimensions, layering consists of stacking two planes with hopping $t_z$ between them; intra-plane hoppings and on-site attraction are the same in both planes. 
We have found that by a judicious choice of fillings and on-site attraction, a bilayer can exhibit critical temperatures between 1.5 and 1.7 times those of the single layer. 
We have also established that double occupancy can be used to distinguish between normal (metallic), 
pre-formed pairs (pseudogap), and superconducting phases through quantum gas microscope measurements.

For a simple cubic lattice, layering consists of allowing $t_z$ to vary with respect to the isotropic case, $t_z=t$, but the number of planes along the $z$ direction, $L_z$, is the same as the number of sites, $L_x=L_y$, along the planar directions.
We have found a smaller enhancement of $T_c$ relative to the isotropic case, namely of 1.3 times larger, for some specific choices of $t_z$, $U$ and $\ave{n}$.

For both geometries, we have found that the BCS estimate roughly tracks the DQMC $T_c$'s, although the positions of $T_c$ maxima are not reliable; less serious is the multiplicative factor of order 20 for both geometries, since the unknown proportionality factor in Eq.\,\eqref{eq:Tc_BCS} could account for it. In addition, since  $T_{c}^{\text{BCS}}$ does not follow the Mermin-Wagner theorem this discrepancy at half-filling in 3D is even more serious for  $t_z\lesssim 0.5t$. Also, care must be taken since the BCS estimate for $T_c$ scales with the bandwidth, which, in turn, depends on $t_z$; this leads to more severe discrepancies in bilayer than in 3D.  

We hope our results will stimulate optical lattice experiments to explore this route of layering to further increase $T_c$.

%%%%%%%%%%%%%%%%%%%%%%%%%%%%%%%%%%%%%%%%%%%%%%%%%%%%%%%%%%%%%%%%%%
%%%%%%%%%%%%%%%%%%%%%%%%%%%%%%%%%%%%%%%%%%%%%%%%%%%%%%%%%%%%%%%%%%
\section*{ACKNOWLEDGMENTS}
%%%%%%%%%%%%%%%%%%%%%%%%%%%%%%%%%%%%%%%%%%%%%%%%%%%%%%%%%%%%%%%%%%
%%%%%%%%%%%%%%%%%%%%%%%%%%%%%%%%%%%%%%%%%%%%%%%%%%%%%%%%%%%%%%%%%%
The authors are grateful to the Brazilian Agencies Conselho Nacional de Desenvolvimento Cient\'\i fico e Tecnol\'ogico (CNPq), Coordena\c c\~ao de Aperfei\c coamento de Pessoal de Ensino Superior (CAPES), and Instituto Nacional de Ci\^encia e Tecnologia de Informa\c c\~ao Qu\^antica (INCT-IQ) for funding this project. 
We also gratefully acknowledge support from Funda\c c\~ao Carlos Chagas de Apoio \`a Pesquisa (FAPERJ), through the grants E-26/200.258/2023 (N.C.C.), E-26/210.974/2024 (R.R.d.S.),  E-26/200.959/2022  
(T.P.), and  E-26/210.100/2023 
(T.P.).
N.C.C. also acknowledges CNPq, Grant No.~313065/2021-7.

%%%%%%%%%%%%%%%%%%%%%%%%%%%%%%%%%%%%%%%%%%%%%%%%%%%%%%%%%%%%%%%%%%%%%%%%
%%%
%%%%%     BIBLIOGRAPHY
%%%%%%%%%%%%%%%%%%%%%%%%%%%%%%%%%%%%%%%%%%%%%%%%%%%%%%%%%%%%%%%%%%%%%%%%%%%

%%%%%%%%%%%%%%%%%%%%%%%%%%%%%%%%%%%%%%%%%%%%%%%%%%%%%%%%%%%%%%%%%%
%%%%%%%%%%%%%%%%%%%%%%%%%%%%%%%%%%%%%%%%%%%%%%%%%%%%%%%%%%%%%%%%%%

\section*{Appendix}
In the context of continuous phase transitions, several quantities display singular behavior in the thermodynamic limit at a critical temperature, $T_c$, characterized by critical exponents. 
For instance, a generic quantity usually diverges as $X\sim|T-T_c|^{-x}$, where $x$ is the associated critical exponent; in particular, the correlation length behaves as $\xi\sim|T-T_c|^{-\nu}$. 
However, in systems with finite linear sizes, $L$, these divergences are rounded, and their limit as $L\to\infty$ is described quite generically by a finite-size scaling (FSS) theory \cite{Fisher71,dosSantos81a,Barber83}, which is based on the fact that the two important length scales are $\xi$ and $L$, so that $X$ follows a scaling ansatz, 
\begin{equation}
	X(L,T) = L^{x/\nu}\mathcal{F}(L/\xi),\ T \rightarrow T_c,\ L\to \infty.
	\label{eq:genFSS}
\end{equation}
In particular, at $T=T_c$ we see that $L^{-x/\nu} X(L,T_c)$ is size-invariant, and could be used to determine $T_c$, if $x$ and $\nu$ were known, which is not usually the case.
However, we may use instead a \emph{dimensionless} quantity, $Y(L,T)$, related to $X$ (so that $y=0$) with which we can estimate $T_c$ without the previous knowledge of $x$ and $\nu$. 
The quest for such quantity begins with the structure factor associated with the relevant order parameter: in the thermodynamic limit, it diverges at a characteristic ordering wavevector, $\qv$, as the critical temperature is approached, i.e., the peak gets narrower and larger.
Thus, with a structure factor, $S(\qv)$, we may form a dimensionless quantity, the so-called correlation ratio, $R_c$, which is scale (or size-) invariant at the critical point, 
\begin{equation}
	R_{c}(L,\beta) = 1 - \frac{S(\mathbf{q}+\delta\mathbf{q})}{S(\mathbf{q})},
	\label{eq:genRc}
\end{equation}	
where $\qv+\delta\mathbf{q}$ is a point in reciprocal space close to $\qv$. 
Equation \eqref{eq:genRc} therefore follows the FSS ansatz, Eq.\,\eqref{eq:genFSS}, with $x=0$, and can then be used to estimate $T_c$ from plots of  $R_{c}(\beta)$ for different $L$'s.
For the case at hand, we use the uniform, $\qv=\mathbf{0}$, $s$-wave pairing structure factor, $P_s(\qv)$, to calculate $R_{c}(L,\beta)$.
\color{black}

\bibliography{ref}
\end{document}